\documentclass[a4paper,11pt]{article}
\pdfoutput=1 

\usepackage{jheppub} 
\usepackage{slashed}
\usepackage{mathtools}
\usepackage{amsfonts}
\usepackage{subcaption}
\usepackage{comment}
\usepackage[T1]{fontenc} 

\title{\boldmath Order parameter and spectral function in $d$-wave holographic superconductors }

 \author{Debabrata Ghorai,}
\author{Taewon Yuk,}
\author{Sang-Jin Sin}
\affiliation{Department of Physics, Hanyang University, Seoul 04763, Korea}

\emailAdd{dghorai123@gmail.com}
\emailAdd{tae1yuk@gmail.com}
\emailAdd{sangjin.sin@gmail.com}

\abstract{
We consider the $d$-wave holographic superconductor model with full backreaction on the metric, addressing a missing part in the literature. We have identified the gap function by comparing the fermionic spectral function with the momentum-dependent order parameter. By numerical investigations of the fermionic spectral function in the presence of a tensor condensate, we find the Fermi arc and the gapped behavior, which closely resemble ARPES data. 
Moreover, we have examined the influence of the coupling constant, chemical potential, and temperature on the spectral function.
We find that  $d$-wave fermionic spectral function can be obtained through  $p_x$ and $p_y$ condensates combined with two fermion flavors. Similarly, combining $d_{x^2-y^2}$ and $d_{xy}$ orbitals symmetry with two fermion flavors leads to a $g$-wave spectral function.

}

\begin{document} 
\maketitle
\flushbottom

\section{Introduction}
ARPES data \cite{RevModPhys.93.025006} indicates that most  unconventional superconductors exhibit $d$-wave orbital symmetry. However, understanding the theoretical aspects of these systems remains elusive due to the limitations of conventional methods in describing strongly coupled systems. To address this, the gauge/gravity duality \cite{Maldacena:1997re, Gubser:1998bc, Witten:1998qj} offers an   approach by employing a weakly coupled dual system in one higher   dimension  \cite{HKMS, Seo:2016vks, Oh:2021xbe,Song:2019asj,Oh:2020cym,Oh:2018wfn,Seo:2017oyh,Seo:2018hrc,Seo:2017yux}.
The relationship between the energy gap   and the critical temperature of high $T_c$ superconductors \cite{Romes:2007} has been given  in \cite{Hart2008}  in the simplest dual gravitational system \cite{Gubser:2008px} with scalar hair. This system exhibits a second-order phase transition from AdS-Schwarzchild geometry to hairy black hole geometry and is referred to as $s$-wave holographic superconductors, characterized by an isotropic energy gap.  
To extend the system   including anisotropic gap function,  we need $p$-wave and $d$-wave  gaps which had been realized in spin one fields   \cite{Gubser:2008wv,Vegh:2010fc,Ghorai:2022gzx}) and tensor fields   \cite{Benini_2010,Chen:2010mk, Zeng:2010fx,Gao:2011aa,Krikun:2013iha,Nishida:2014lta,Krikun:2015tga} respectively. 

\noindent   Considerable  amount of works  \cite{Horo2009,Horowitz:2009ij,Horo2011,Ammon:2010pg,Gubser:2010dm,Sin:2009wi,Brihaye:2010mr,Siop2010,Zeng:2010vp,Siopsis:2011vs,zeng2011analytical,Pan:2012jf,Siop2012,gangopadhyay2012analytic,Kim:2013oba,Cai_2014,DeWolfe:2016rxk,ghorai2016higher,Ghorai:2016tvk,Srivastav:2019tkr,Ghorai:2021uby,Donini:2021kne} have been done  using  gravitons and photons in the bulk. Although  less attention has been paid to the fermionic side, there have been some works on the fermion spectral function, exhibiting distinct spectral features in the presence of the scalar \cite{Faulkner:2009am,Yuk:2022lof}, vector\cite{Vegh:2010fc,Ghorai:2023wpu}, and tensor \cite{Benini_2010,Chen:2010mk, Benini:2010qc,Chen:2011ny} condensations. 
The presence of some of such condensations  gives rise to the    Fermi arc for $p$ and $d$-wave holographic superconductors.
In the case  of spin two fields, the Lagrangian density becomes somewhat intricate. The initial  formulation of   $d$-wave holographic superconductivity \cite{Chen:2010mk} did not  treat  the   number of degrees of freedom properly. 
 Based on earlier investigations on the spin two fields \cite{Buchbinder:1999ar,Buchbinder:2000fy}, the formulation of an  proper action for a massive charged spin two field was accomplished by Benini, Herzog, Rahman, and Yarom (BHRY) in \cite{Benini_2010}, by employing the Einstein condition which forbids the back reaction, so that in their setup we have to  investigate $d$-wave holographic superconductors in the probe limit only. However, the full backreacted geometry  \cite{Ge:2012vp,Kim:2013oba,Xu:2022qjm} can play an important role \cite{Hartnoll:2008kx}.  \\

\noindent In this paper, we reformulate the BHRY Lagrangian by replacing the Einstein condition with the traceless condition in the constraint equation for spin-two fields. 
This allows the presence of the backreaction of tensor condensate to the metric. Another concern in  $d$-wave holographic superconductors is about the momentum dependence of the order parameter, which should be consistent with that of the fermion spectral function. In all previous computations, the order parameter was considered as $B_{xx}$, which is independent of the momentum direction.   One of our aims here is to identify the precise $d$-wave order parameter that has angular dependence in momentum space consistent  with that of the fermion spectral function. 
We will identify  the correct $d$-wave order parameter as $B_{\rho\rho}$ where $\rho$ is the radial direction in the $xy$ plane.
The detailed analysis of the Fermi arc  in the presence of the  $d$-wave gap and full backreaction is presented here for various orbital symmetries. 
We also examine the  effect of coupling strength, the chemical potential and temperature on the spectral function. We have shown that the $d$-wave spectral function can be obtained from the two different $p$-wave condensates with two fermion flavors. Similarly, two fermion flavors with condensates of two different tensor fields  lead to $g$-wave spectral function. 
These   may be useful to describe higher orbital superconductivity.\\~

\noindent This paper is organized as follows. In Section 2,  we argue that one can construct a Lagrangian density without imposing the Einstein condition. This allows  the presence of the back-reaction of matter fields on the background geometry. In Section 3, we numerically investigate the critical temperature and all bosonic configurations.
In Section 4, we  describe how to calculate the boundary fermion Green's function using the flow equation. In Section 5, we analyze the spectral function  of various cases. 
We summarize and conclude in Section 6.

\section{Basic set up for spin two field} 
The Fierz-Pauli Lagrangian for a spin-2 field in flat space, presented in \cite{Fierz:1939ix}, is given by
\begin{eqnarray}
\mathcal{L}= \frac{1}{4}\left[-\partial_{\rho}B_{\mu\nu}\partial^{\rho}B^{\mu\nu}+ 2 B_{\mu} B^{\mu}- 2 B^{\mu}\partial_{\mu}B^{\alpha}_{~\alpha}+\partial_{\mu}B^{\alpha}_{~\alpha} \partial^{\mu}B^{\alpha}_{~\alpha}-m^2 \left(B_{\mu\nu}B^{\mu\nu} -B^{\alpha}_{~\alpha}B^{\alpha}_{~\alpha} \right)  \right] ~~~~~~
\end{eqnarray}
where $B_{\mu}=\partial^{\alpha}B_{\alpha\mu}$. The corresponding equation of motion along with the constraints is 
\begin{eqnarray}
\partial^{\alpha}\partial_{\alpha}B_{\mu\nu}-m^2 B_{\mu\nu} &=& 0 \\
\partial_{\mu}B^{\mu\nu} = 0 ~~~&\text{and}&~~~  B^{\mu}_{~\mu} = 0 ~. 
\end{eqnarray}
This yields the correct number of degrees of freedom for the dynamics of a spin-2 field in flat space. The Lagrangian  construction for a neutral massive spin-2 field  with   correct counting of the   degrees of freedom was developed in \cite{Buchbinder:2000fy} using two distinct methods.
In one approach, the  condition  of vanishing Einstein tensor, $G_{\mu\nu}=0$, was employed. In this scenario, all matter fields can not exert a backreaction on the metric. In the alternate method, the static background condition was adopted to ensure the consistency of constraint equations, allowing matter fields to affect the metric. A detailed analysis of the Reissner-Nordstrom black hole solution was presented in \cite{Buchbinder:2000fy}.
We closely follow  the procedure  in \cite{Buchbinder:2000fy} where the Lagrangian for a charged massive spin-2 field in $AdS$ spacetime was formulated under the Einstein condition, as described in \cite{Benini_2010}. We argue that the Einstein condition can be replaced by the traceless condition of the field, which is one of the constraints equation of the field.
Then the backreaction of matter fields   can be considered.
The Lagrangian for a symmetric tensor in curved space can be expressed as follows \cite{Benini_2010}:
\begin{eqnarray}
	\mathcal{L}= -|D_{\rho}B_{\mu\nu}|^2 + 2 |D_{\mu}B^{\mu\nu}|^2+ |D_{\mu}B^{\rho}_{~\rho}|^2-\left(D_{\mu}B^{*\mu\nu}D_{\nu}B^{\rho}_{~\rho}+ h.c. \right) -m^2 \left( |B_{\mu\nu}|^2-|B^{\mu}_{~\mu}|^2\right) \nonumber \\ + c_1 R_{\mu\nu\rho\lambda} B^{*\mu\rho} B^{\nu\lambda} +c_2 R_{\mu\nu} B^{*\mu\rho} B^{\nu}_{\rho} + c_3 R |B_{\mu\nu}|^2 + iq c_4 F_{\mu\nu}B^{*\mu\rho} B^{\nu}_{\rho}+ c_5 R |B^{\rho}_{~\rho}|^2 \nonumber \\
	+ c_6 \left(e^{i\phi} R_{\mu\nu}B^{*\mu\nu}B^{\rho}_{~\rho} + h.c. \right) ~~~~
\end{eqnarray} 
where $B_{\mu}=D^{\alpha}B_{\alpha\mu}$ and $R_{\mu\nu\rho\lambda}, R_{\mu\nu}, R$ are the Riemann tensor, Ricci curvature and Ricci scalar of the background spacetime respectively. 
The corresponding equation of motion
\begin{eqnarray}
E_{\mu\nu} &=& \left( D^{\alpha}D_{\alpha}- m^2\right) B_{\mu\nu} -(D_{\mu}B_{\nu}+D_{\nu}B_{\mu})+ \frac{1}{2}\left(D_{\mu}D_{\nu}B^{\rho}_{~\rho}+D_{\nu}D_{\mu}B^{\rho}_{~\rho} \right) + g_{\mu\nu}D^{\alpha}B_{\alpha}  \nonumber \\ && -g_{\mu\nu}\left(D^{\alpha}D_{\alpha}-m^2\right) B^{\rho}_{~\rho} + c_1 R_{\mu\rho\nu\lambda} B^{\rho\lambda} +\frac{c_2}{2} (R_{\mu \alpha} B^{\alpha}_{\nu}+R_{\nu \alpha} B^{\alpha}_{\mu}) + c_3 R B_{\mu\nu} \nonumber \\ && + c_4 \frac{iq}{2} (F_{\mu\alpha} B^{\alpha}_{\nu}+F_{\nu\alpha} B^{\alpha}_{\mu}) + c_5 g_{\mu\nu} R  B^{\rho}_{~\rho}+ c_6 \left(e^{i\phi} R_{\mu\nu} B^{\rho}_{~\rho}+ e^{-i\phi}g_{\mu\nu}R_{\alpha\beta}B^{\alpha\beta} \right)  ~~~~~~
\end{eqnarray}
We now try to calculate all the coefficient $c_{i}$ from the constraint equation $D^{\mu}D^{\nu}E_{\mu\nu}=0$ which gives \cite{Benini_2010}
\begin{eqnarray}
(c_1 -2) R^{\mu\alpha\nu\beta}D_{\mu}D_{\nu}B_{\alpha\beta} + c_2 R^{\mu\alpha} D_{\mu}B_{\alpha}+ (c_3 R-m^2) D^{\alpha}B_{\alpha} + iq (1+c_4) F^{\alpha\beta}D_{\alpha}B_{\beta} \nonumber \\
+ \left(c_5 R + m^2 \right)D^{\mu}D_{\mu}B^{\rho}_{~\rho}+ c_6 e^{-i\phi} R^{\alpha\beta}D^{\mu}D_{\mu}B_{\alpha\beta}+\left(1+ c_6 e^{i\phi}\right)R^{\mu\alpha}D_{\mu}D_{\alpha}B^{\rho}_{~\rho}+...=0 ~~~~~
\end{eqnarray}
where the ellipsis denotes terms which containts at most single derivative of spin 2 field. Since no second derivative term exists in any constraint equation, all terms involved in second-order derivatives in the above equation must vanish. This leads to contradictory equations for $c_6$. To remove this contradiction, we use the traceless condition of the spin two field $(B^{\rho}_{~\rho}=0)$ instead of the Einstein condition. Substituting $B^{\rho}_{~\rho}=0$ in the above equation, we get
\begin{eqnarray}
(c_1 -2) R^{\mu\alpha\nu\beta}D_{\mu}D_{\nu}B_{\alpha\beta} + c_2 R^{\mu\alpha} D_{\mu}B_{\alpha}+ (c_3 R-m^2) D^{\alpha}B_{\alpha}+ iq (1+c_4) F^{\alpha\beta}D_{\alpha}B_{\beta} \nonumber \\  + c_6 e^{-i\phi} R^{\alpha\beta}D^{\mu}D_{\mu}B_{\alpha\beta} +...=0
\end{eqnarray}
The above equation becomes a constraint equation when all second derivative terms are eliminated, which leads   
\begin{eqnarray}
c_1=2, ~c_2=0, ~c_3=0, ~c_4=-1, ~c_6=0  ~~\text{and}~~ m^2=0 ~~.
\end{eqnarray}
In the above method, we have not used Einstein condition to determine   $c_i$. Therefore, we can consider the backreaction of the matter field on the background geometry.
The Lagrangian for a traceless symmetric spin two field with Maxwell's term then is given by
\begin{eqnarray}
\mathcal{L}_m = - |D_{\alpha}B_{\mu\nu}|^2+2 |D_{\mu}B^{\mu\nu}|^2 + 2 R_{\mu\nu\rho\lambda}B^{*\mu\rho}B^{\nu\lambda}-iq F_{\mu\nu}B^{*\mu\lambda}B^{\nu}_{\lambda}  -\frac{1}{4}F_{\mu\nu}F^{\mu\nu} ~.
\end{eqnarray}

\section{Holographic $d$-wave superconductor}
The action for $d$-wave holographic superconductors reads
\begin{eqnarray}
S= \frac{1}{2\kappa^2}\int d^{4}x \sqrt{-g} \left[R- 2 \Lambda+ 2\kappa^2 \mathcal{L}_m\right]
\end{eqnarray}
where $\Lambda$ is the cosmological constant and $\kappa^2$ is the gravitational Newton constant. 
The equation of motions for the tensor field and gauge field yield
\begin{eqnarray}
\Box B_{\alpha\beta}- \left(D_{\alpha}B_{\beta} + D_{\beta}B_{\alpha}\right) + 2 R_{\alpha\mu\beta\nu}B^{\mu\nu} -\frac{iq}{2}(F_{\alpha\mu}B^{\mu}_{\beta}+F_{\beta\mu}B^{\mu}_{\alpha}) &=& 0 ~~\\
D_{\mu} F^{\mu\nu}-  \left[iq B^{*}_{\alpha\beta}\left(D^{\nu} B^{\alpha\beta} - D^{\alpha} B^{\nu\beta} \right) + iq B^{*}_{\alpha}B^{\nu\alpha} + h.c.\right] &=& 0
\end{eqnarray}
The Einstein field reads
\begin{eqnarray}
R_{\alpha\beta} -\frac{1}{2} g_{\alpha\beta} R + \Lambda g_{\alpha\beta} = 2 \kappa^2 T_{\alpha\beta}
\end{eqnarray}
where $T_{\alpha\beta} = \frac{1}{2} g_{\alpha\beta}\mathcal{L}_m - \frac{\delta \mathcal{L}_m}{\delta g^{\alpha\beta}}$. Considering the backreacted four-dimensional metric in the following simplified form:
\begin{eqnarray}
ds^2= \frac{L^2}{z^2}\left[ -f(z) g(z) dt^2 + \frac{dz^2}{f(z)} + dx^2 + dy^2 \right] 
\end{eqnarray}
where $L$ is the $AdS$ radius. For this given black hole geometry, the Hawking temperature is
\begin{eqnarray}
 T_H=\frac{|f'(z_h)|\sqrt{g(z_h)}}{4\pi}.
\end{eqnarray}
To introduce both symmetry $d_{x^2-y^2}$ and $d_{xy}$ in the system, we consider the tensor field in following form: 
\begin{eqnarray}
	B= \frac{L^2 \phi(z)}{\sqrt{2}z^2} \left[\alpha \left( dx^2 - dy^2 \right)+ 2\beta dxdy  \right]
	\label{tanstz}
\end{eqnarray}
When $\alpha=0$ ($\beta=0$), we will get only $d_{xy}$ ($d_{x^2-y^2}$) symmetry. 
The matter field ansatz in general reads
\begin{eqnarray}
	B = B_{xx}(z) dx^2 + B_{xy}(z)dxdy +B_{yx}(z)dydx+ B_{yy}(z)dy^2
\end{eqnarray}
Using the symmetric and traceless condition of the spin two field, the above matter field ansatz in polar coordinate become
\begin{eqnarray}
	B= B_{\rho\rho} d\rho^2 +2 B_{\rho\theta}d\rho d\theta + B_{\theta\theta}d\theta^2 ~~~~
\end{eqnarray}
where 
\begin{eqnarray}
	B_{\rho\rho} = \cos2\theta B_{xx}+ \sin2\theta B_{xy}~  &,&
	\frac{B_{\theta\theta}}{\rho^2}= - \cos2\theta B_{xx}+ \sin2\theta B_{xy}~~, \nonumber\\ \frac{B_{\rho\theta}}{\rho} &=& - \sin2\theta B_{xx} + \cos2\theta B_{xy}
\end{eqnarray}
Since $B_{\theta\theta}$ and $B_{\rho\theta}$ have coordinate singularity, the angle dependent order parameter for $d$-wave superconductor is $B_{\rho\rho}$. 
With the tensor field ansatz (\ref{tanstz}), the corrected order parameter reads
\begin{eqnarray}
	B_{\rho\rho}= \frac{L^2 \phi(z)}{\sqrt{2}z^2} \left[\alpha \cos2\theta + \beta \sin2\theta \right] ~~.
	\label{orderp}
\end{eqnarray}

\noindent With gauge field ansatz $A=A_{t}(z) dt$, equation of motion of all fields are
\begin{eqnarray}
	\label{eomf1}
g'(z)+ 2 \kappa^2 z d_m^2 \left[g(z) \phi '(z)^2 + \frac{2g(z) \phi (z)^2}{z^2} +\frac{ q^2 A_t(z)^2 \phi (z)^2}{f(z)^2}\right] &=& 0~~~ \\
\label{eomf2}
f'(z) -\frac{3 f(z)}{z}+\frac{3}{z}+\frac{f(z) g'(z)}{2 g(z)} + \frac{2 \kappa^2}{z}d_m^2 \left[ f(z) \phi(z)^2 -\frac{ z^4 A_t'(z)^2}{4L^2 d_m^2 g(z)}\right] &=& 0~~~~~~~~\\
A_t''(z)-\frac{g'(z)}{2 g(z)}A_t'(z)-\frac{2q^2 L^2 d_m^2 \phi (z)^2 }{z^2 f(z)}A_t(z) &=& 0 \\ 
\phi''(z)+\left[\frac{f'(z)}{f(z)}+ \frac{g'(z)}{2 g(z)}-\frac{2 }{z} \right]\phi '(z) +\left[\frac{q^2 A_t(z)^2 }{f(z)^2 g(z)} \right]\phi (z) &=& 0
	\label{eomf4}
\end{eqnarray}
where $d_m^2=|\alpha|^2 +|\beta|^2$. 
The horizon condition $f(z_h)=0$ and eq.(\ref{eomf2}) lead to determine the Hawking temperature which reads
    \begin{eqnarray}
    	T_H = \frac{3\sqrt{g(z_h)}}{4\pi z_h} \left[1-\frac{\kappa^2}{3}\left(\frac{z_h^3A_t^{\prime}(z_h)^2}{2L^2 g(z_h)} \right) \right]  ~.
    	\label{hawkingtemp2}
    \end{eqnarray}
    \noindent At the boundary of this spacetime should be asymptotically AdS spacetime which imposes the boundary condition on $g(z=0)=1$. Therefore the field equations at the boundary becomes
    \begin{eqnarray}
    A^{\prime\prime}(z) =0 ~~~~~\text{and}~~~~~~~~ \phi^{\prime\prime}(z) -\frac{2}{z}\phi'(z)=0 ~.
    \end{eqnarray} 
    which gives the asymptotic behaviour of the gauge field and vector field in terms of quantities of boundary theory in following way: 
    \begin{eqnarray}
    	A_t(z) = \mu -\tilde{\rho} z ~~~\text{and} ~~~ \phi(z) = C_s + C_c z^{3}
    \end{eqnarray}
    where $\mu, \tilde{\rho}, C_s, C_c $ are the chemical potential, charge density, source term and expection value of angle independent condensation of the boundary theory respectively.
    From eq.(\ref{orderp}),  the angle dependent order parameter for $d$-wave holographic superconductor can be mapped with the condensation value of the boundary theory in following way: 
    \begin{eqnarray}
    B_{\rho\rho}=\frac{L^2 \phi(z)}{\sqrt{2}z^2} \left[\alpha \cos2\theta + \beta \sin2\theta \right] = \langle\mathcal{O}\rangle z
    \end{eqnarray}
     when the source ($C_s$) is zero. Therefore, the angle dependent condensation operator reads 
    \begin{eqnarray}
    \langle\mathcal{O}\rangle=  \frac{L^2 C_c}{\sqrt{2}} \left[\alpha \cos2\theta + \beta \sin2\theta \right] = \frac{L^2 C_c l}{\sqrt{2}} \cos(2\theta- \theta_1)
    \label{condval}
    \end{eqnarray}
    where $l^2=\alpha^2+\beta^2$ and $\tan\theta_1=\frac{\beta}{\alpha}$ are determined from the given value of real value $\alpha$ and $\beta$. The above expression clearly shows that the mixing of $d_{x^2-y^2}$ and $d_{xy}$ symmetry in system rotates the gap structure. In order to determine the angle dependent condensation operator value, we now need to calculate the $C_c$ value in the full backreacted system. This gives us momentum dependent gap structure in Fourier space $(k_x, k_y)$ with help of  the following identification:
    \begin{eqnarray}
    	 \Delta_k = FT[\langle\mathcal{O}\rangle]= \frac{1}{2\pi a^2}\int_{0}^{a} \int_{0}^{2\pi} \langle\mathcal{O}\rangle  e^{-i (k_x \rho\cos\theta + k_y \rho\sin\theta)} \rho d\rho d\theta.
    	 \label{2dfourier}
    \end{eqnarray} 
     where $a$ is the sample size and $FT[...]$ is the two dimensional Fourier transformation. 
  
	\subsection{The critical temperature and momentum dependent order parameter}
     In this subsection, we will employ the Shooting method to numerically solve the system of coupled equations, given by equations (\ref{eomf1})-(\ref{eomf4}). With help of scaling symmetry \cite{Chen:2011ny,Xu:2022qjm} of the equation of motion of all fields,  we can set $L=1$ and $2\kappa^2=1$. To successfully solve these equations, it is essential to provide appropriate boundary conditions for all fields, which are
     \begin{align}
     	A_{t}(z_h)=0, ~~~~~f(z_h)=0,  ~~~~~g(0)=1, ~~~~~C_s=0 ~.
     \end{align} 
     By specifying $(T, \mu)$ parameters, we are able to obtain solutions for all the equations in the system. To unveil the near-horizon behavior of the fields, we employ a Taylor series expansion, allowing us to express the fields as follows:
	\begin{eqnarray}
		\label{hexpansion}
		(V_y(z), A_t(z), f(z), g(z)) \approx \sum_{i=0}^{5} (V_{yi}, A_{ti}, f_{i}, g_{i})\left(1-\frac{z}{z_h}\right)^{i}
		\label{pbeom11}
	\end{eqnarray}
    By substituting the aforementioned expansion into the field equations, as given by equations (\ref{eomf1})-(\ref{eomf4}), we establish a relation between the coefficients $(V_{yi}, A_{ti}, f_{i}, g_{i})$ and the horizon data $(V_{y0}, A_{t1}, z_h, g_{0})$. Through the imposition of the boundary conditions and the subsequent solution of the equations of motion for the fields, we can determine the horizon data for a given combination of $(T, \mu)$, denoted as $(T_0, \mu_0)$. The utilization of these field equations, as presented in equation (\ref{pbeom11}), in conjunction with the solution for the horizon data, yields the complete configurations of all fields for a desired ratio $\frac{T}{\mu}$.
    
\begin{figure}[h!]
   	\centering
   	\begin{subfigure}[b]{0.32\textwidth}
   		\centering
   		\includegraphics[scale=0.22]{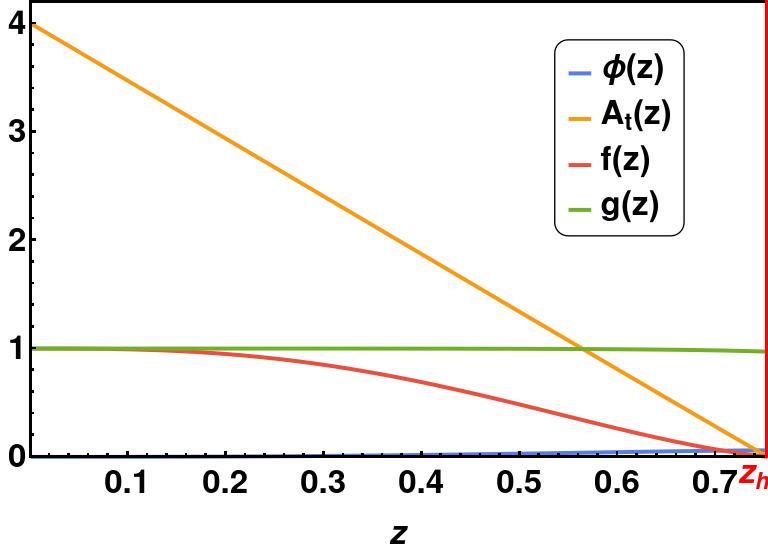}
   		\caption{$T=0.975 T_c $}
   	\end{subfigure}
   	\hfil
   	\begin{subfigure}[b]{0.32\textwidth}
   		\centering
   		\includegraphics[scale=0.22]{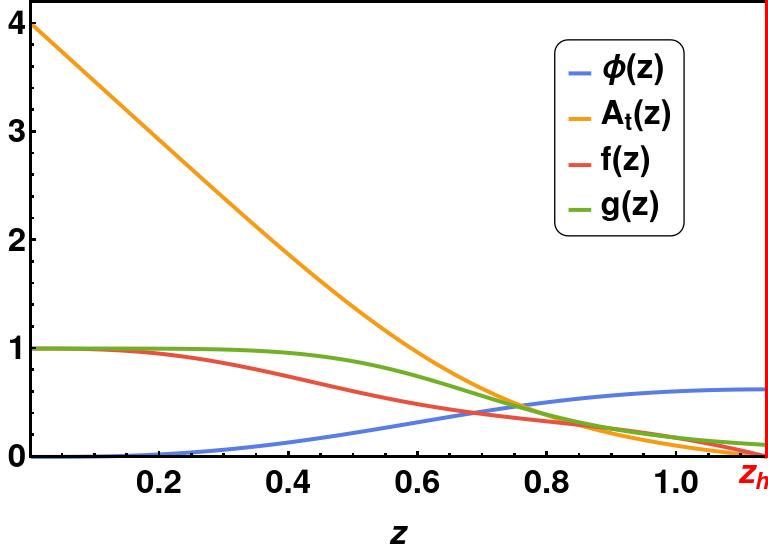}
   		\caption{$T=0.5 T_c $}
   	\end{subfigure}
   	\hfil
   	\begin{subfigure}[b]{0.32\textwidth}
   		\centering
   		\includegraphics[scale=0.22]{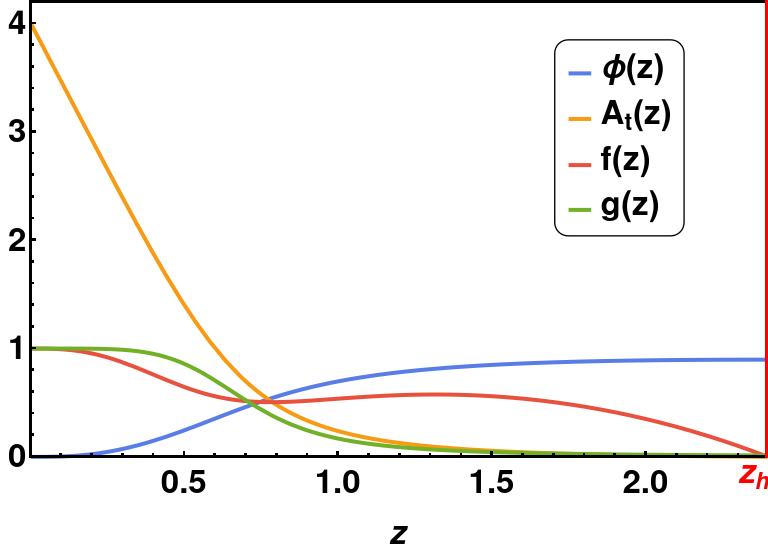}
   		\caption{$T=0.125 T_c$}
   	\end{subfigure}
   	\caption{Backreacted profiles at three different temperatures.}
   	\label{fig1bm}
   \end{figure}

\begin{figure}[h!]
 	\centering
 	\begin{subfigure}[b]{0.3\textwidth}
 		\centering
 		\includegraphics[scale=0.25]{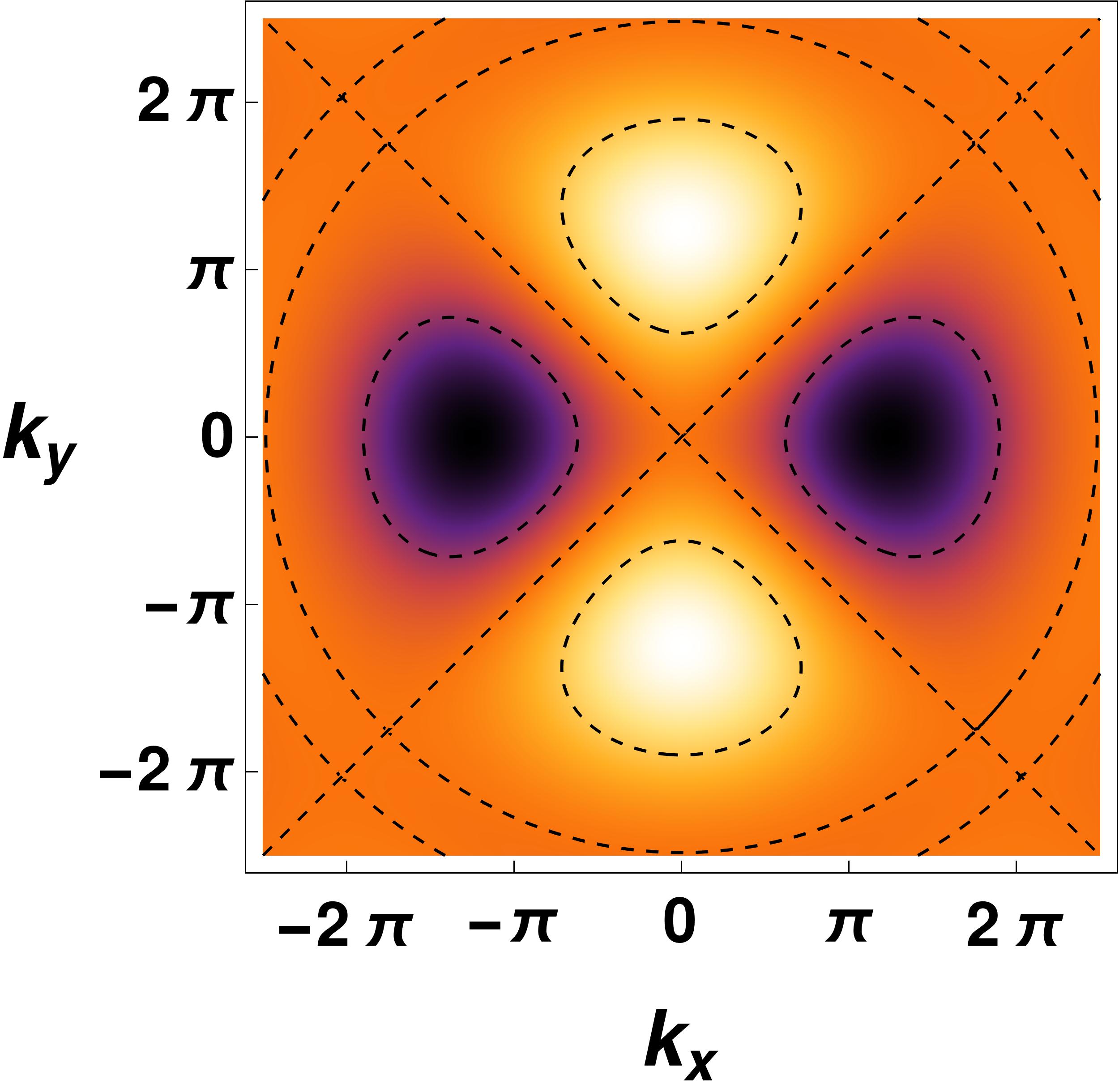}
 		\caption{$d_{x^2-y^2}$-wave}
 	\end{subfigure}
 	\hfil
 	\begin{subfigure}[b]{0.3\textwidth}
 		\centering
 		\includegraphics[scale=0.25]{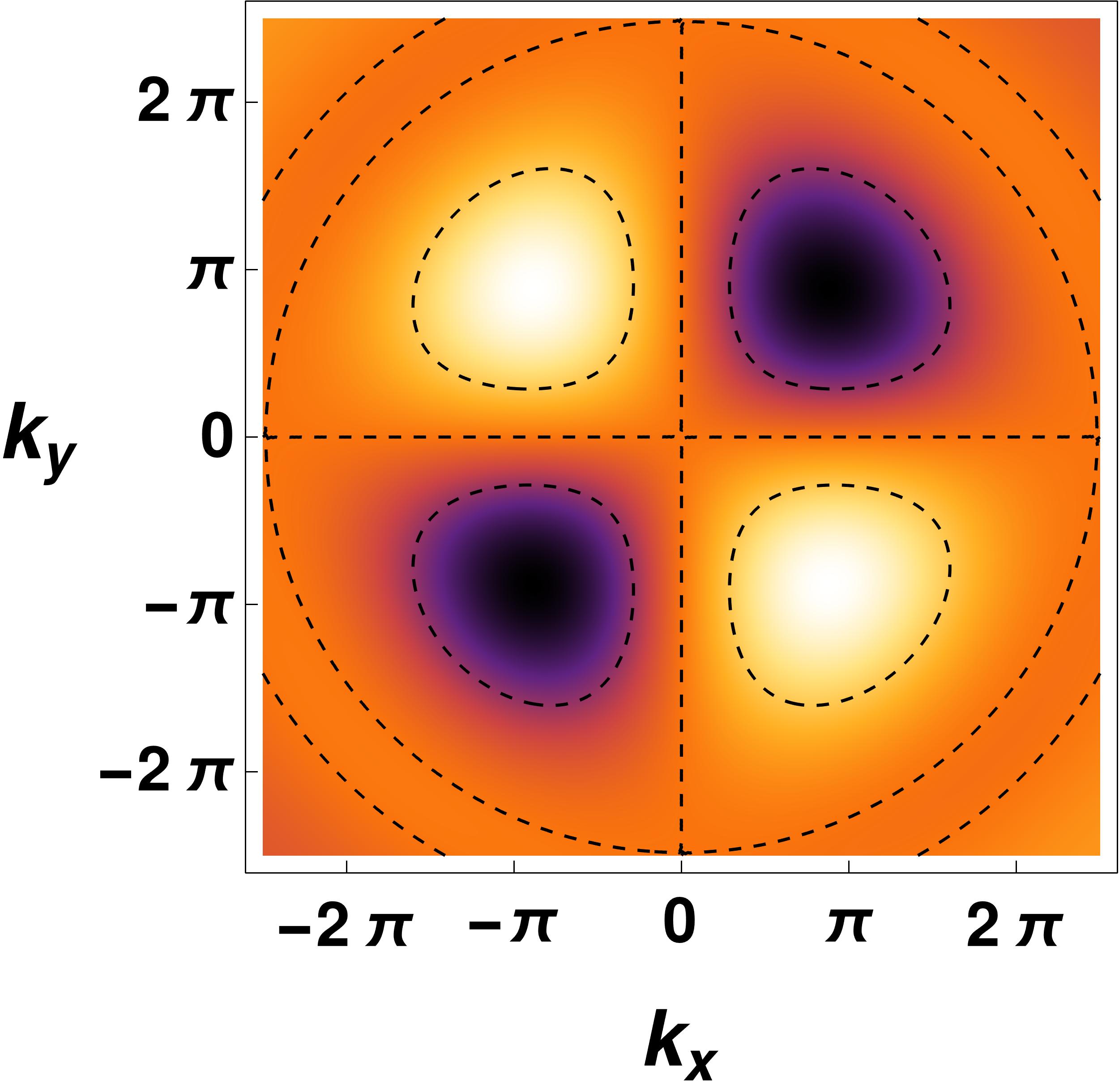}
 		\caption{$d_{xy}$-wave }
 	\end{subfigure}
 	\hfil
 	\begin{subfigure}[b]{0.3\textwidth}
 		\centering
 		\includegraphics[scale=0.25]{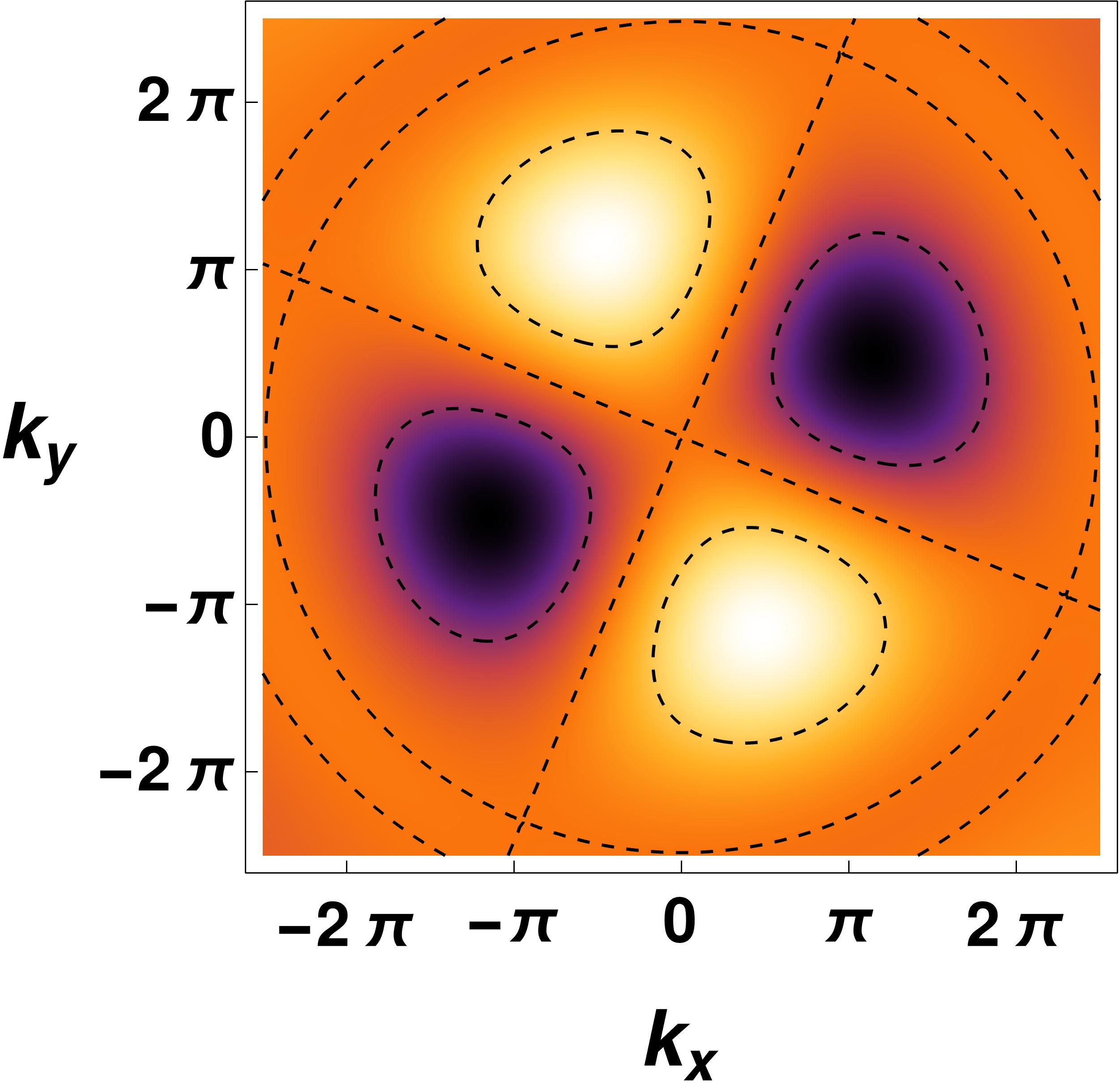}
 		\caption{$d_{x^2-y^2}+d_{xy}$-wave}
 	\end{subfigure}
 	\caption{Momentum dependent order parameter at $T=0.125 T_c $ for different symmetries.}
 		\label{figgapm}
 \end{figure}

\noindent   With the choice of the charge of the tensor field $q=2$, and utilizing equation (\ref{hawkingtemp2}), we have determined that $T_c \approx 0.02 \mu$. Remarkably, this critical temperature remains consistent regardless of the symmetry parameter, whether it is for $d_{xy}$-wave $(\alpha=0, \beta=1)$ superconductivity, or for $d_{x^2-y^2}$-wave $(\alpha=1, \beta=0)$ superconductivity, or for $d_{x^2-y^2}+d_{xy}$-wave $(\alpha=\frac{1}{\sqrt{2}}, \beta=\frac{1}{\sqrt{2}})$ superconductivity. Furthermore, for a fixed temperature, the condensation value is approximately $C_c \approx 0.05522 \mu^3$ for all three types of superconductivity: $d_{xy}$-wave, $d_{x^2-y^2}$-wave, and $d_{x^2-y^2}+d_{xy}$-wave.
   The field configurations are illustrated in Figure \ref{fig1bm}. 
   Moving forward, we employ this field solution and equation (\ref{2dfourier}) to plot the momentum-dependent gap structure at $T=0.125 T_c$, as depicted in Figure \ref{figgapm}. Subsequently, in a subsequent section, we will explore the fermionic spectral function, which will exhibit a Fermi arc at a $45^\circ$ angle in momentum space for the $d_{x^2-y^2}$-wave symmetry. This analysis will provide support for the proposition that the order parameter in $d$-wave holographic superconductors should be $B_{\rho\rho}$ instead of $B_{xy}$ or $B_{xx}$.

\section{Fermion with tensor condensation}
\subsection{Fermionic set up}
	The fermionic action can be expressed as \cite{Benini_2010}:
	\begin{eqnarray}
	S_{\psi} &=& \int d^4x \sqrt{-g} \left[i \bar{\psi} (\Gamma^{\mu}D_{\mu}- m_{f})\psi -i \bar{\psi}_c (\Gamma^{\mu}D^{*}_{\mu}- m_{f})\psi_c +\mathcal{L}_{int} \right]\nonumber \\
    \mathcal{L}_{int} &=& \eta^{*} B^{*}_{\mu\nu} \bar{\psi_c} \Gamma^{\mu}D^{\nu}\psi- \eta \bar{\psi}\Gamma^{\mu} D^{\nu}(B_{\mu\nu} \psi_{c}) ~~.
	\end{eqnarray}
	Here, $\eta$ represents the coupling constant. The spinor's covariant derivative is denoted by $D_{\mu}=\partial_{\mu} + \frac{1}{4} \omega_{\mu\bar{\alpha\beta}}\Gamma^{\bar{\alpha\beta}}- i q_f A_{\mu}$. Additionally, the field $\psi_{c}=\psi^{*}$ corresponds to the complex charge conjugate field of the fermion, which is treated as an independent field throughout the computations in this framework.
	The boundary action for standard quantization \cite{Iqbal:2009fd,Laia:2011zn} is given by:
	\begin{eqnarray}
     	S_{bdy}= i \int d^3x  \sqrt{-h} (\bar{\psi}\psi+ \bar{\psi}_c \psi_{ c}) ~.
	\end{eqnarray}
	For this formulation, we adopt the following set of bulk gamma matrices:
	\begin{eqnarray}
		\Gamma^{\underline{t}} = \sigma_1 \otimes i\sigma_2, ~~ \Gamma^{\underline{x}} = \sigma_1 \otimes \sigma_1, ~~ \Gamma^{\underline{y}} = \sigma_1 \otimes \sigma_3, ~~
		\Gamma^{\underline{z}} = \sigma_3 \otimes \sigma_0 
	\end{eqnarray} 
    where underline indices represent tangent space indices.
    We obtain the Dirac equation \cite{Benini_2010}
    \begin{eqnarray}
	(\Gamma^{\mu}D_{\mu}- m_{f})\psi &+& i \mathcal{I}_{int} \psi_c = 0   
    \end{eqnarray}
    where $ \mathcal{I}_{int}= 2  \eta B_{\mu\nu} \Gamma^{\mu}D^{\nu} +  \eta B_{\mu}\Gamma^{\mu} $. 
    To simplify the analysis, we express the fermionic field as follows:
    \begin{eqnarray}
	\psi (t, x, y, z) = \frac{1}{(-gg^{zz})^{1/4}} e^{-i \omega t + i k_x x + i k_y y } \Psi(z) ~~.
	\label{eq26}
   \end{eqnarray}
    This form allows us to eliminate the spin connection term in the spinor's equation of motion. By substituting the aforementioned spinor into the Dirac equations, we derive the following expressions:
    \begin{eqnarray}
	\left[\Gamma^{\underline{z}}\partial_z - i \left( \sqrt{\frac{g^{tt}}{g^{zz}}}(\omega+ q_f A_t)\Gamma^{\underline{t}} - \sqrt{\frac{g^{xx}}{g^{zz}}} k_x \Gamma^{\underline{x}} -  \sqrt{\frac{g^{yy}}{g^{zz}}} k_y \Gamma^{\underline{y}}\right) -\frac{m_f}{\sqrt{g^{zz}}} \right] \Psi (z)   \nonumber \\
	 + \frac{i}{\sqrt{g^{zz}}}\tilde{\mathcal{I}}_{int} \Psi_c (z) = 0 
    \label{eq27a1}
   \end{eqnarray} 
    where $\tilde{\mathcal{I}}_{int}= - \frac{\sqrt{2}i\eta\phi(z)}{z^2} (g^{xx})^{\frac{3}{2}}\left[ \alpha \left(k_x \Gamma^{\underline{x}}- k_y \Gamma^{\underline{y}}\right)+ \beta \left(k_x \Gamma^{\underline{y}}+ k_y \Gamma^{\underline{x}} \right) \right] $ since the metric is isotropic $xy$ plane. The field equation for conjugate fermion is 
    \begin{eqnarray}
	\left[\Gamma^{\underline{z}}\partial_z + i \left( \sqrt{\frac{g^{tt}}{g^{zz}}}(\omega - q_f A_t)\Gamma^{\underline{t}} - \sqrt{\frac{g^{xx}}{g^{zz}}} k_x \Gamma^{\underline{x}} -  \sqrt{\frac{g^{yy}}{g^{zz}}} k_y \Gamma^{\underline{y}}\right) -\frac{m_f}{\sqrt{g^{zz}}} \right] \Psi_c (z) \nonumber \\
	 - \frac{i}{\sqrt{g^{zz}}}\tilde{\mathcal{I}}_{int} \Psi (z) = 0
    \label{eq27b1}
    \end{eqnarray} 
    We express the four-component spinor as
    \begin{eqnarray}
	\Psi(z)= \begin{pmatrix}
		\Psi_{+}(z) \\
		\Psi_{-}(z)
	\end{pmatrix}, ~~~~~~\text{where}~~ \Psi_{\pm}= \begin{pmatrix}
		\Psi_{\pm 1} \\
		\Psi_{\pm 2}
	\end{pmatrix}
\end{eqnarray}
which allows us to formulate the Dirac equation as follows
\begin{eqnarray}
	\left[\partial_z \mp \frac{m_f}{\sqrt{g^{zz}}}\right] \Psi_{\pm}(z) =\pm  \left[ i K_{\mu}\gamma^{\mu}\Psi_{\mp}(z) + \frac{\sqrt{2}\eta g^{xx}\phi(z)}{z^2} \left(\alpha K_{\mu} \gamma^{\mu}_{(\alpha)} + \beta K_{\mu} \gamma^{\mu}_{(\beta)}  \right) \Psi_{c\mp}(z)\right] ~~~~~
	\label{eq39}
\end{eqnarray}
where $K_{\mu}= \left(\sqrt{\frac{g^{tt}}{g^{zz}}} (\omega +q_f A_t), -\sqrt{\frac{g^{xx}}{g^{zz}}}k_x, -\sqrt{\frac{g^{yy}}{g^{zz}}}k_y\right)$, $\gamma^{\mu}=\left(i\sigma_2, \sigma_1, \sigma_3\right), \gamma^{\mu}_{(\alpha)}=\left(0, \sigma_1, -\sigma_3\right) $ and $\gamma^{\mu}_{(\beta)}=\left(0, \sigma_3, \sigma_1\right)$. 
In a similar manner, we can reformulate the equation of motion for the conjugate fermion. In the asymptotic limit as $z\rightarrow 0$, we consider $g^{\mu\nu}\rightarrow z^2 \eta^{\mu\nu}$, where $\eta^{\mu\nu}$ is the Minkowski metric. Consequently, the behavior of the spinor in this regime is given by:
\begin{eqnarray}
	\Psi_{+} (z) &=& \mathbf{A} z^{m_f} + \mathbf{B} z^{1-{m_f}},  ~~~~~ \Psi_{-}(z) = \mathbf{D} z^{-{m_f}} + \mathbf{C} z^{1+{m_f}} \\
	\Psi_{c+} (z) &=& \tilde{\mathbf{A}}^{*} z^{m_f} + \tilde{\mathbf{B}}^{*} z^{1-{m_f}},  ~~~~~ \Psi_{c-}(z) = \tilde{\mathbf{D}}^{*} z^{-{m_f}} + \tilde{\mathbf{C}}^{*} z^{1+{m_f}}
\end{eqnarray}
Here, $\mathbf{A}, \mathbf{B}, \mathbf{C}, \mathbf{D}$ are two-component spinors that are determined by solving the complete bulk Dirac equations. For $|{m_f}|<\frac{1}{2}$, the leading term yields the boundary spinor solutions as follows:
\begin{eqnarray}
	\Psi (z) \approx \begin{pmatrix}
		\mathbf{A} z^{m_f} \\
		\mathbf{D} z^{-{m_f}} 
	\end{pmatrix},  ~~~~~\Psi_c (z) \approx \begin{pmatrix}
		\tilde{\mathbf{A}}^{*} z^{m_f} \\
		\tilde{\mathbf{D}}^{*} z^{-{m_f}} 
	\end{pmatrix}
	\label{eq211a}
\end{eqnarray}
Remarkably, we have found that the leading order of the asymptotic behavior of the fields is always $z^{\pm m_f}$ for $|m_f|<\frac{1}{2}$, and this behavior remains independent of the interaction. The two-component spinor captures the influence of the interactions.
Following same procedure as in \cite{Ghorai:2023wpu}, we can write down the boundary action in following form
\begin{eqnarray}
	S_{bdy} = \int d^3x  \left[ \xi^{(C)\dagger} \tilde{\Gamma}\xi^{(S)}+ \xi^{(S)\dagger} \tilde{\Gamma}\xi^{(C)} \right]
	\label{eq218}
\end{eqnarray}
where the boundary gamma matrix $\tilde{\Gamma}= \sigma_0 \otimes (-\sigma_2) $ and the source and condensation are given by 
\begin{eqnarray}
  \xi^{(S)}= \begin{pmatrix} 
  	\Psi_{+} \\
  	\Psi_{c-}
  \end{pmatrix} \overset{z\rightarrow 0}{=}\begin{pmatrix}
  \mathbf{A} z^{m_f} \\
  \tilde{\mathbf{D}}^{*} z^{-{m_f}}
\end{pmatrix},  ~~~~~~\text{and}~~ \xi^{(C)}= \begin{pmatrix} 
  \Psi_{-} \\
  \Psi_{c+}
  \end{pmatrix} \overset{z\rightarrow 0}{=}  \begin{pmatrix}
  \mathbf{D} z^{-{m_f}} \\
  \tilde{\mathbf{A}}^{*} z^{{m_f}}
\end{pmatrix} ~~.
	\label{eq319}
\end{eqnarray}
This is the Namubu-Gorkov spinor representation, which represents the particle-hole symmetry.

\subsection{Green function from flow equation}
Rearranging all components of eq.(s)(\ref{eq27a1},\ref{eq27b1}), we can recast the Dirac equations in following structure
\begin{eqnarray}
	\label{eq215}
	\partial_z \xi^{(S)} + \mathbb{M}_1 \xi^{(S)}  + \mathbb{M}_2 \xi^{(C)} &=& 0 \\
	\partial_z \xi^{(C)} + \mathbb{M}_3 \xi^{(C)}  + \mathbb{M}_4 \xi^{(S)} &=& 0 
	\label{eq216}
\end{eqnarray}
where $4\times4$-matrix $\mathbb{M}_{i}, ~i=1,2,3,4$ are determined from (\ref{eq27a1},\ref{eq27b1}). We have calculated those $\mathbb{M}_{i}$ which are 
\begin{eqnarray}
	\mathbb{M}_1= \begin{pmatrix}
		\mathbb{N}_1 & \mathbb{P}_1 \\
		-\mathbb{P}_1 & - \mathbb{N}_1
	\end{pmatrix}, 		~~~\mathbb{M}_2= \begin{pmatrix}
		\mathbb{N}_2(q) & 0 \\
		0 & \mathbb{N}_2 (-q)
	\end{pmatrix}, 	~~~\mathbb{M}_3= -\mathbb{M}_1, ~~~~ \mathbb{M}_4= -\mathbb{M}_2 ~~~~~~~
\end{eqnarray}
where
\begin{eqnarray}
	\mathbb{N}_1 = - \frac{m_f}{z \sqrt{f(z)}}\boldsymbol{1}_{2\times 2}, ~~~~	\mathbb{P}_1 =\frac{\sqrt{2}\eta \phi(z)}{\sqrt{f(z)}}\left[\alpha \begin{pmatrix}
		- k_y &  k_x  \\
		 k_x  &  k_y
	\end{pmatrix} + \beta \begin{pmatrix}
	 k_x &  k_y \\
	  k_y &  -k_x
\end{pmatrix} \right]  \nonumber \\
\mathbb{N}_2 (q) = \frac{i}{\sqrt{f(z)}} \begin{pmatrix}
	k_y  & k_x-\frac{(\omega+ q A_t(z))}{\sqrt{f(z) g(z)}}\\
	k_x+\frac{(\omega+ q A_t(z))}{\sqrt{f(z) g(z)}} & -k_y
\end{pmatrix}	~~~
\end{eqnarray}
Following the procedure in \cite{Yuk:2022lof,Ghorai:2023wpu}, we get flow equation of bulk Green's function in following form
\begin{eqnarray}
	\partial_z \mathbb{G}(z) + \tilde{\Gamma} \mathbb{M}_3 \tilde{\Gamma} \mathbb{G}(z) - \mathbb{G}(z) \mathbb{M}_1 -\mathbb{G}(z) \mathbb{M}_2\tilde{\Gamma} \mathbb{G}(z)+ \tilde{\Gamma} \mathbb{M}_4 =0
	\label{equflowm}
\end{eqnarray}
From the horizon behaviour of spinor, we can find the horizon value of the bulk Green's function \cite{Yuk:2022lof}
\begin{eqnarray}
	\mathbb{G} (z_h) = i \bold{1}_{4\times 4} ~~.
	\label{eqgreenha}
\end{eqnarray}
 The boundary retarded Green's function is determined the solution of bulk Green's function at the boundary as follow
\begin{eqnarray}
	\mathbb{G}_{r} = \lim_{z\to 0} U(z) \mathbb{G}(z) U(z) 
\end{eqnarray}
where $U(z)= diag(z^m, z^m, z^{-m}, z^{-m})$ and $\mathbb{G}_r$ is the retarded Green function, defined from the boundary action \cite{Iqbal:2009fd}. For the numerical evaluation of the Green function, we will fix the mass of the fermion to be zero ($m_f = 0$) and charge of fermion to be one since $q=2q_f$. The fermionic spectral function is defined as 
\begin{eqnarray}
	A(\omega, k_x, k_y)= Tr[Im [\mathbb{G}_{r}]] ~~.
\end{eqnarray}
In the presence of fully backreacted bosonic fields, we will study the spectral function and will compare the spectral function in probe limit case with backreaction case in the next section.

\section{Fermionic gap in the presence of tensor condensation}
 
 \noindent To incorporate the Fermi arc characteristic into holographic superconductors, we need to examine the fermionic spectral function within the framework of holographic superconductors. To acquire the fermionic spectral function, we employ numerical methods to solve the flow equation (\ref{equflowm}), utilizing the bulk Green function (\ref{eqgreenha}) evaluated at the horizon. In the probe limit, we examine the fermionic spectral function in the AdS-Schwarzschild background, wherein the influence of the bosonic matter field on the underlying spacetime is disregarded. 
 In the absence of any bosonic condensate, the holographic fermion setup unveils a Fermi surface, whose Fermi momentum is dictated by the chemical potential. The emergence of the gap feature necessitates the introduction of interactions between the fermion field and a bosonic field.
 
\begin{figure}[h!]
	\centering
	\includegraphics[scale=0.22]{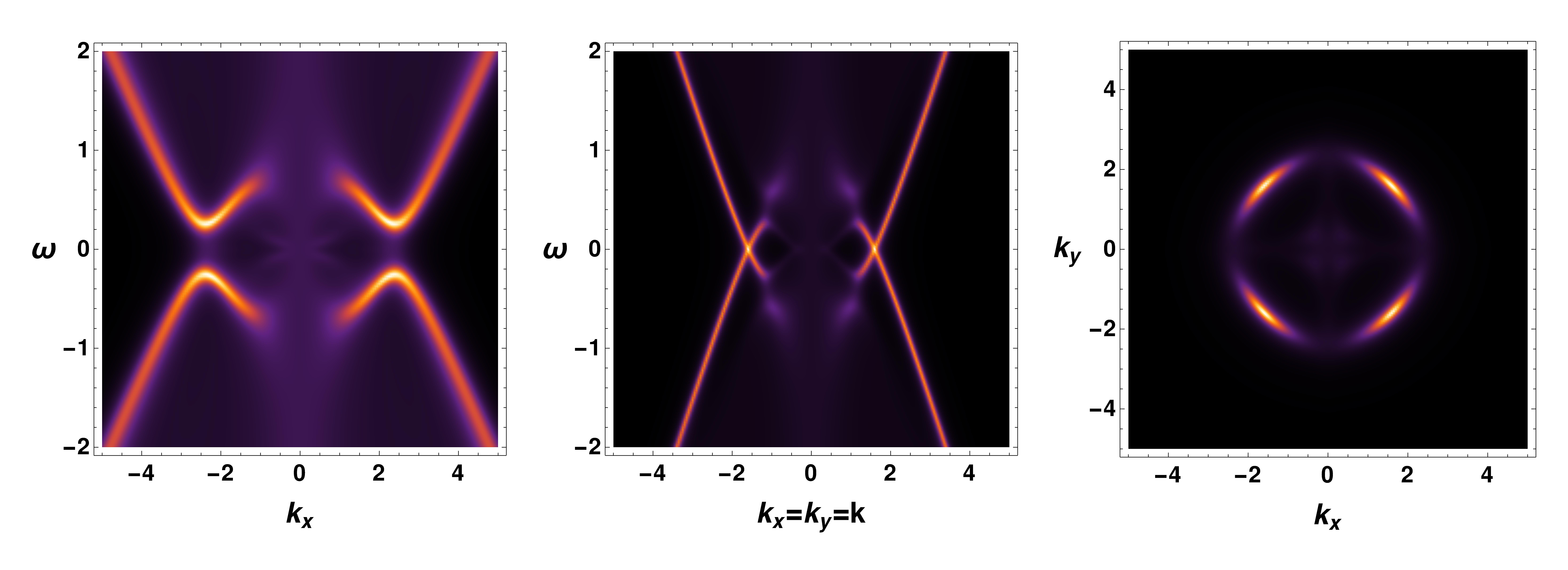}
	\caption{Spectral function for $d_{x^2-y^2}$-condensate $(\alpha=1, \beta=0,\eta=1)$ at $ T=0.125 T_c$.}
	\label{FigSF2}
\end{figure}

\begin{figure}[h!]
	\centering
	\begin{subfigure}[b]{0.24\textwidth}
		\centering
		\includegraphics[scale=0.21]{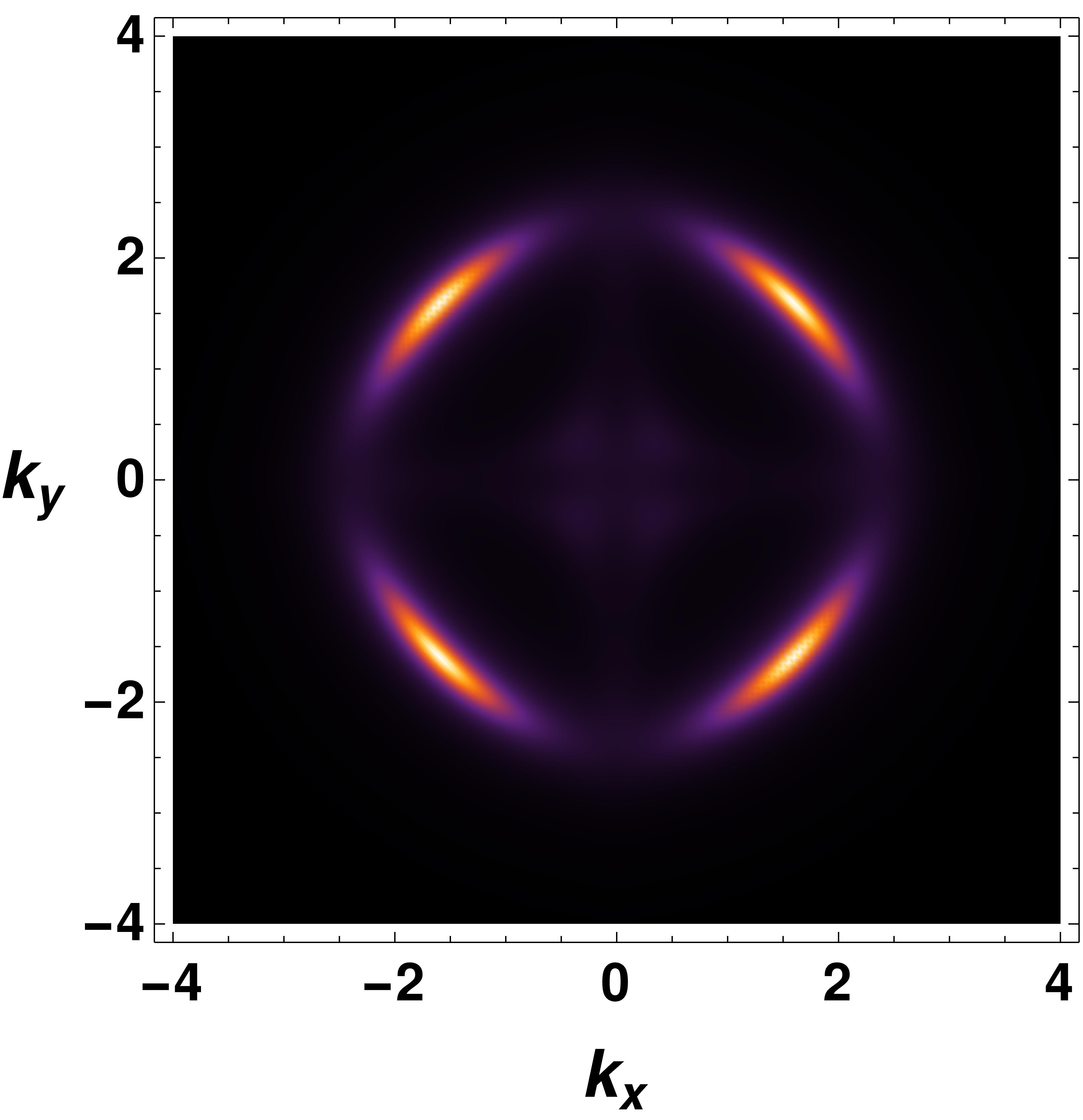}
		\caption{$d_{x^2-y^2}$-wave: \\
			$~~~~~(\alpha,\beta)=(1,0)$}
	\end{subfigure}
	\hfil
	\begin{subfigure}[b]{0.24\textwidth}
		\centering
		\includegraphics[scale=0.21]{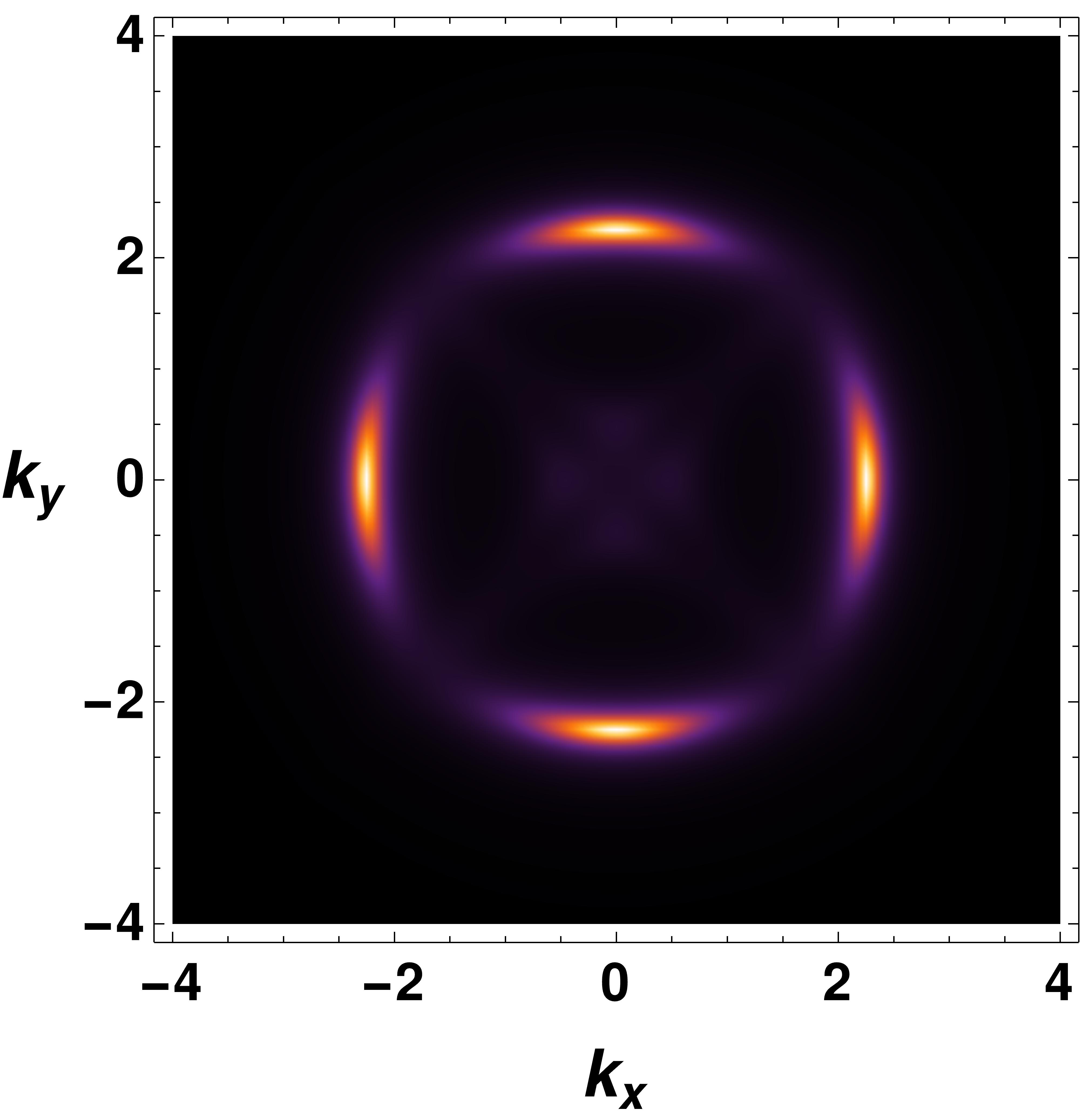}
		\caption{$d_{xy}$-wave: \\
			$~~~~~(\alpha,\beta)=(0,1)$}
	\end{subfigure}
	\hfil
	\begin{subfigure}[b]{0.24\textwidth}
		\centering
		\includegraphics[scale=0.21]{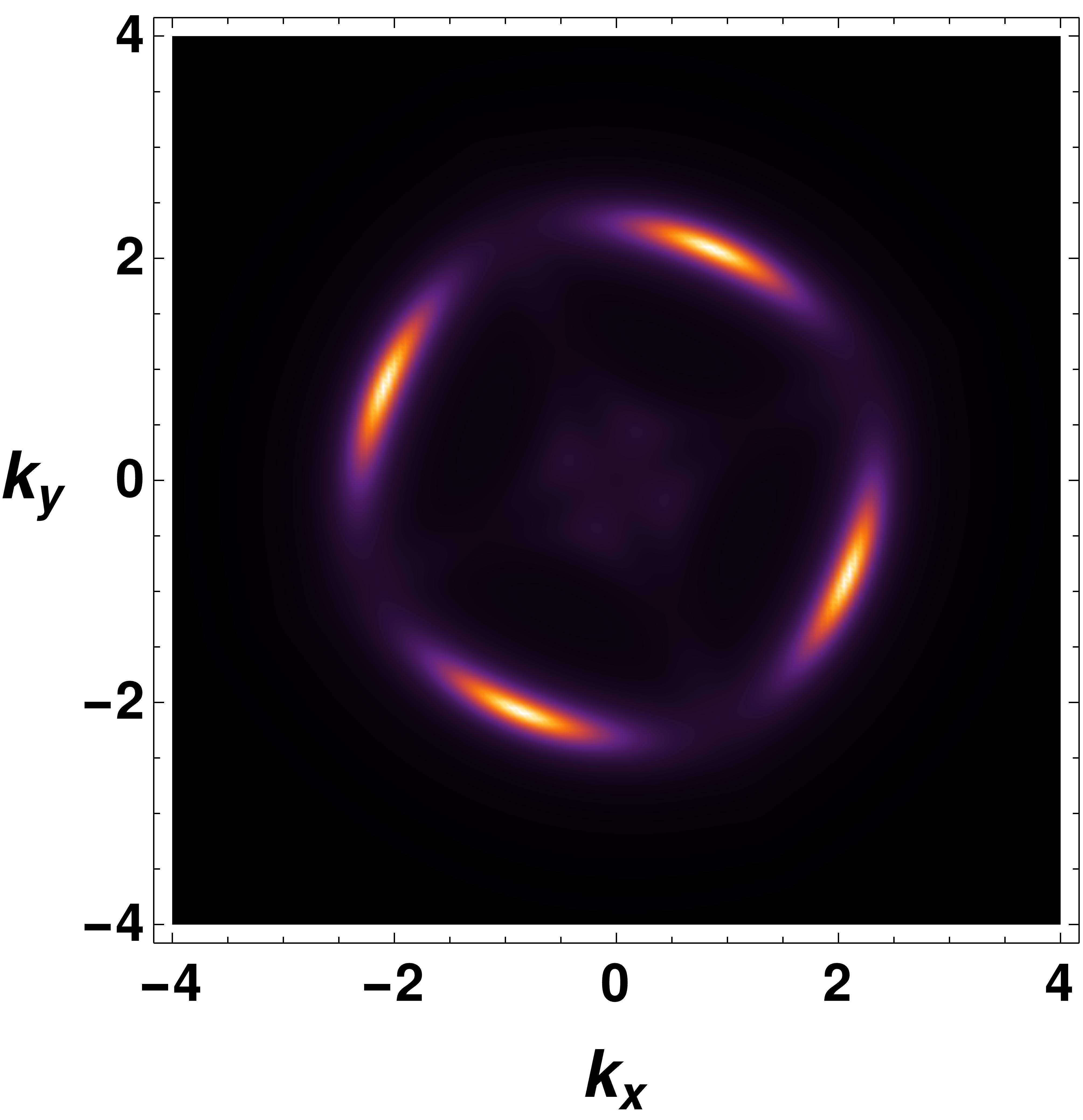}
		\caption{$d_{x^2-y^2}+d_{xy}$-wave:\\
			$~~~~~~(\alpha,\beta)=(\frac{1}{\sqrt{2}},\frac{1}{\sqrt{2}})$}
	\end{subfigure}
		\hfil
	\begin{subfigure}[b]{0.24\textwidth}
		\centering
		\includegraphics[scale=0.21]{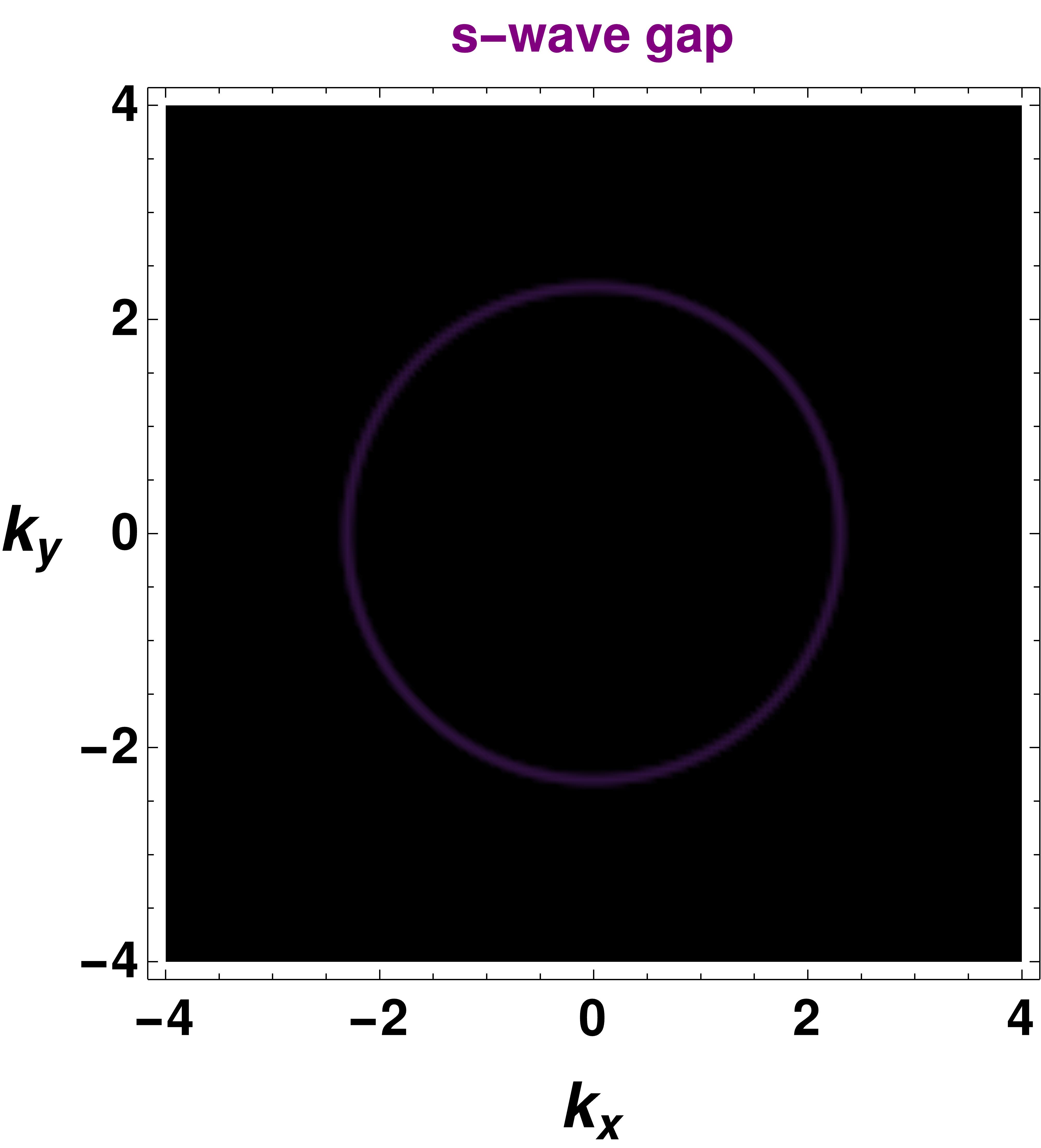}
		\caption{$d_{x^2-y^2}+i d_{xy}$-wave:\\
		$~~~~~~(\alpha,\beta)=(\frac{1}{\sqrt{2}},\frac{i}{\sqrt{2}})$}
	\end{subfigure}
	\caption{Fermi arc in spectral function at $T=0.125 T_c$ with different $d$-wave symmetries. The $d_{x^2-y^2}+i d_{xy}$-wave gives $s$-wave fermionic spectral function.}
	\label{comdx2dxyb}
\end{figure}

\begin{figure}[h!]
	\centering		
	\begin{subfigure}[b]{0.45\textwidth}
		\centering
		\includegraphics[scale=0.21]{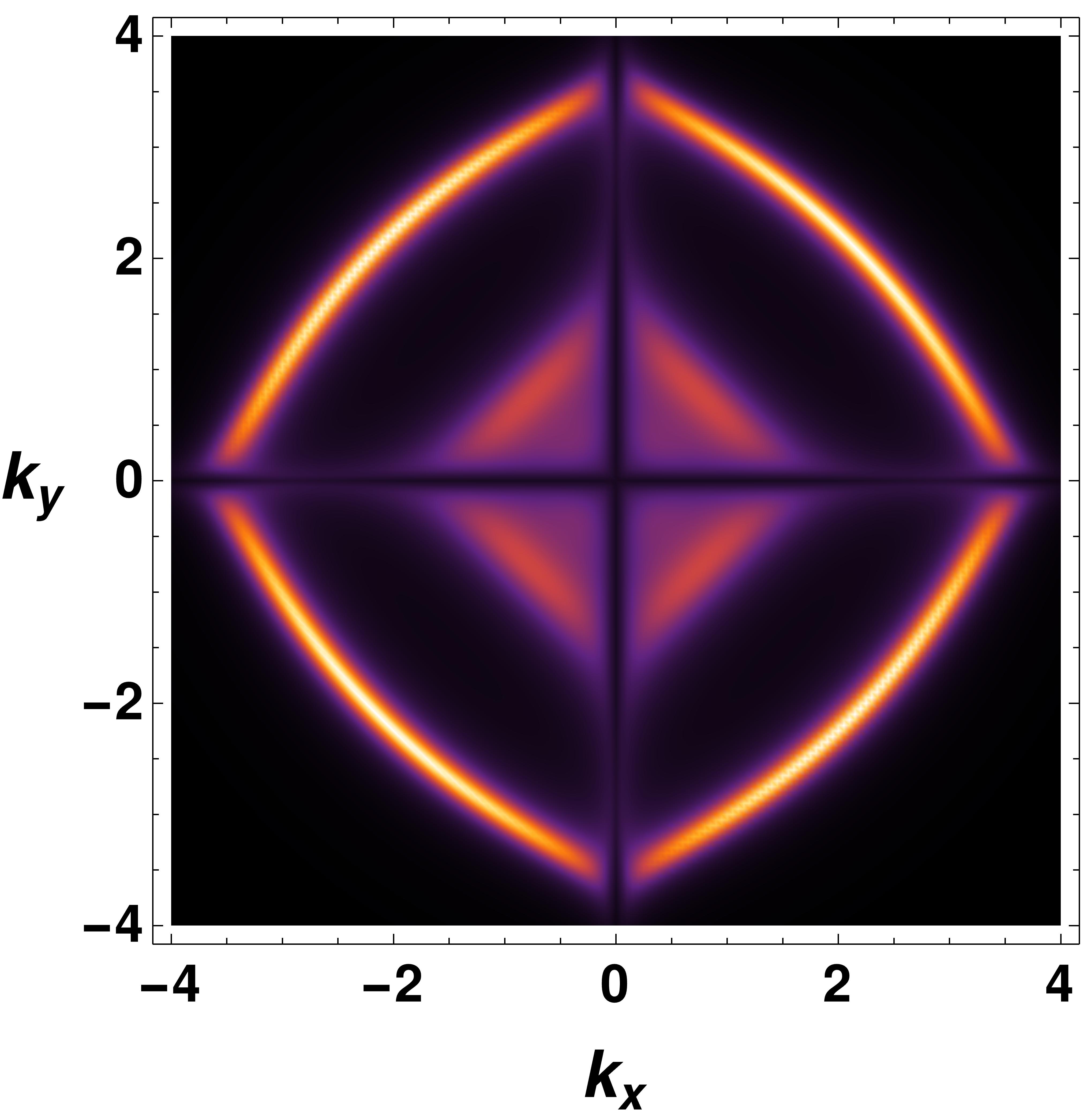}
		\caption{Probe limit}
	\end{subfigure}
	\hfil
	\begin{subfigure}[b]{0.45\textwidth}
		\centering
		\includegraphics[scale=0.21]{FigD/SFdx2T01mu4XYeta1.jpeg}
		\caption{Backreaction}
	\end{subfigure}
	\caption{Spectral function in probe limit case and backreacted cases at $T=0.125 T_c$.}
	\label{FigSF3}
\end{figure}

\begin{figure}[h!]
	\centering		
	\begin{subfigure}[b]{0.3\textwidth}
		\centering
		\includegraphics[scale=0.2]{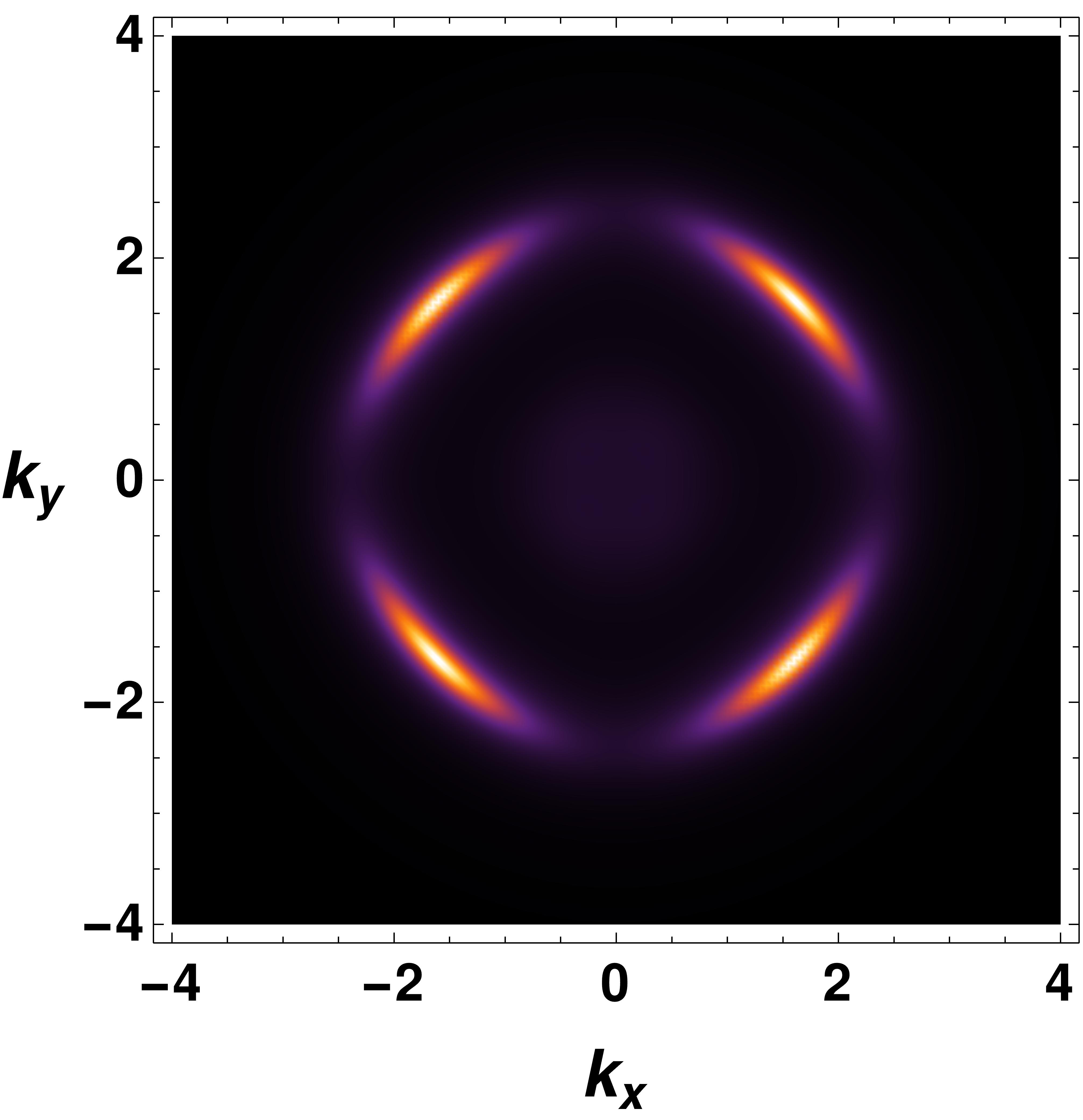}
		\caption{$T=0.5 T_c $}
	\end{subfigure}
	\hfil
	\begin{subfigure}[b]{0.3\textwidth}
		\centering
		\includegraphics[scale=0.2]{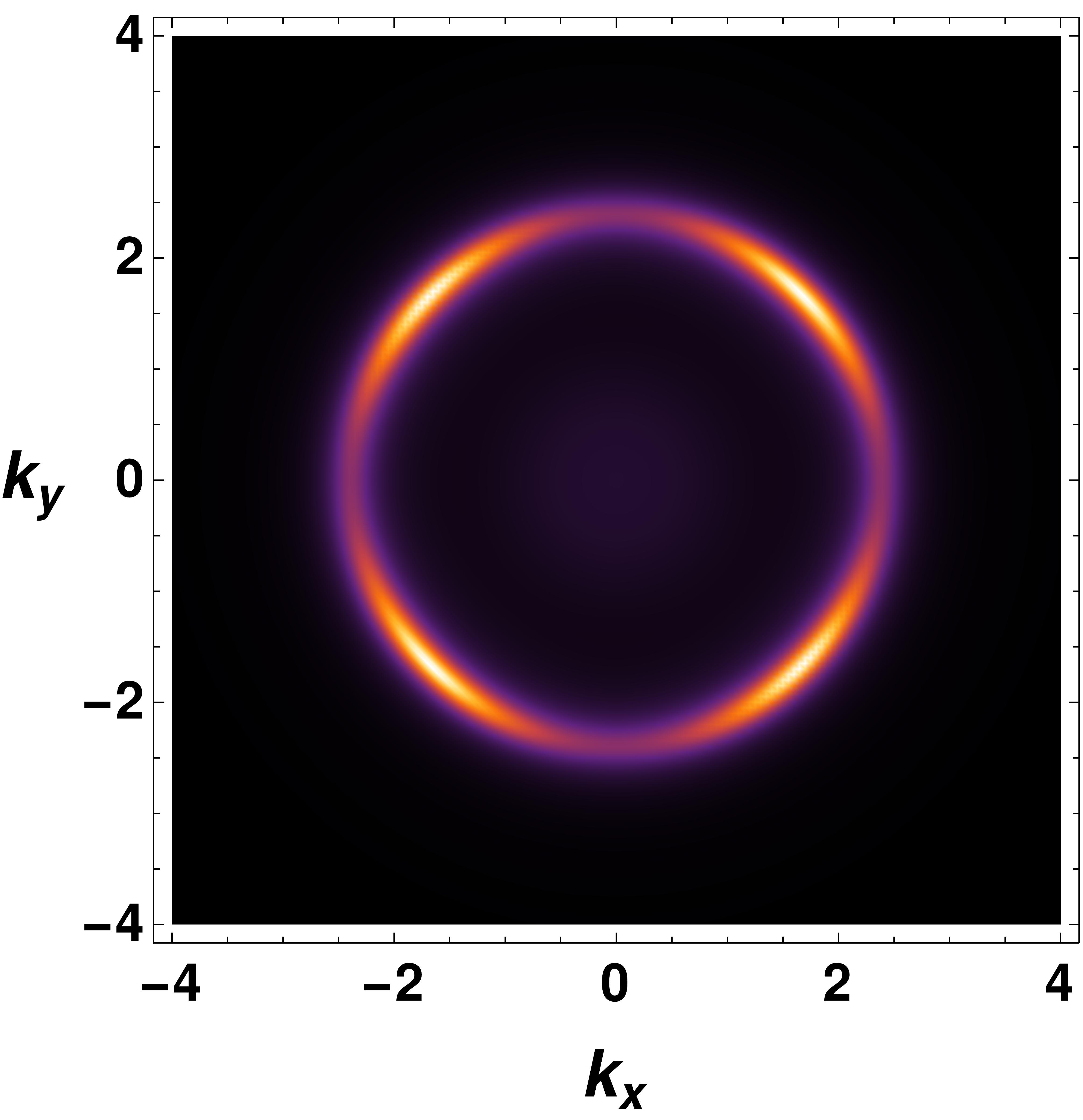}
		\caption{$T=0.875 T_c$}
	\end{subfigure}
	\hfil
	\begin{subfigure}[b]{0.3\textwidth}
		\centering
		\includegraphics[scale=0.2]{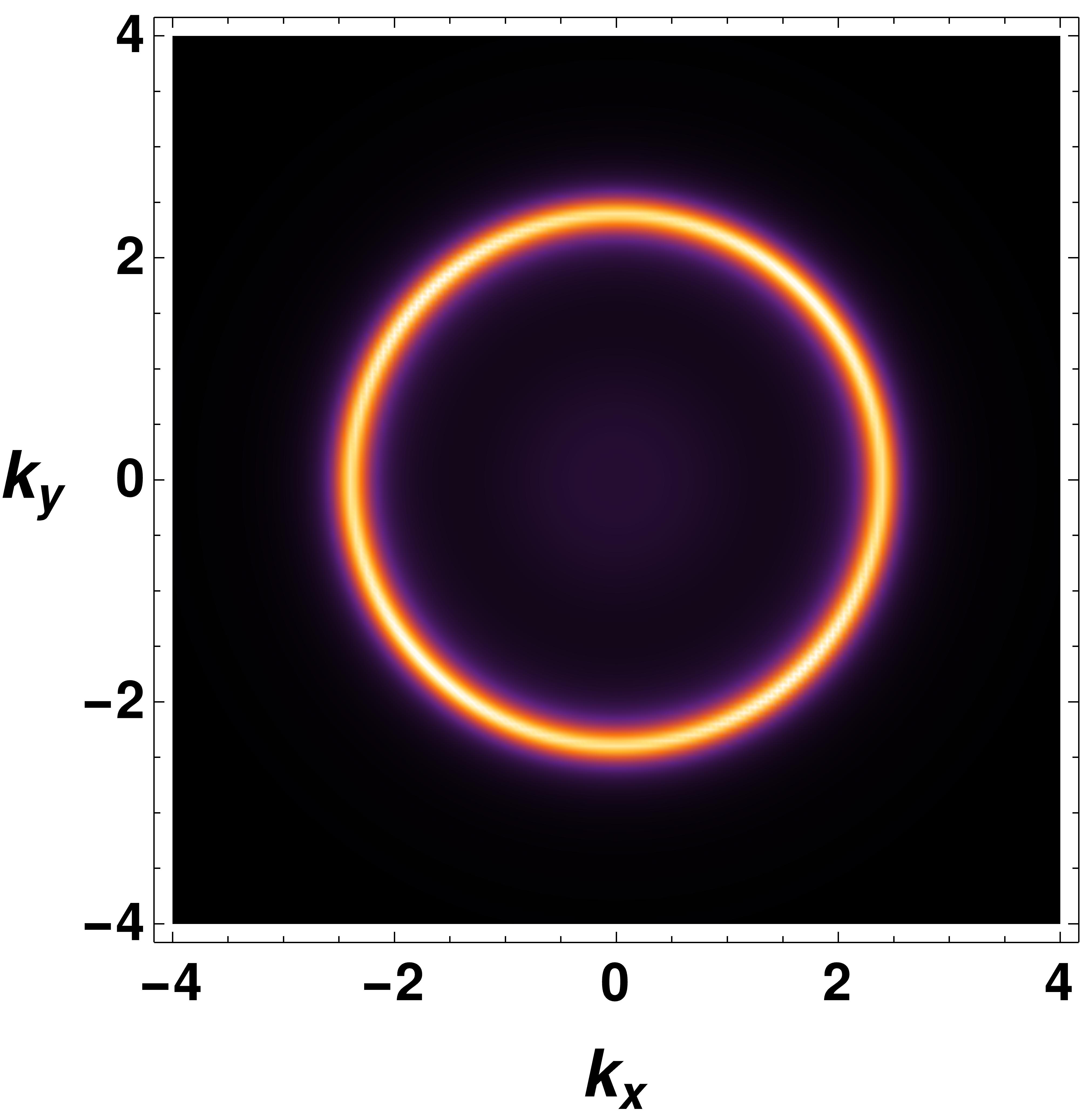}
		\caption{$T \geq T_c $}
	\end{subfigure}
	\caption{Effect of temperature on spectral function with $d_{x^2-y^2}$-condensate for $\eta=1$.}
	\label{FigSF3t}
\end{figure}

\begin{figure}[h!]
	\centering		
	\begin{subfigure}[b]{0.3\textwidth}
		\centering
		\includegraphics[scale=0.23]{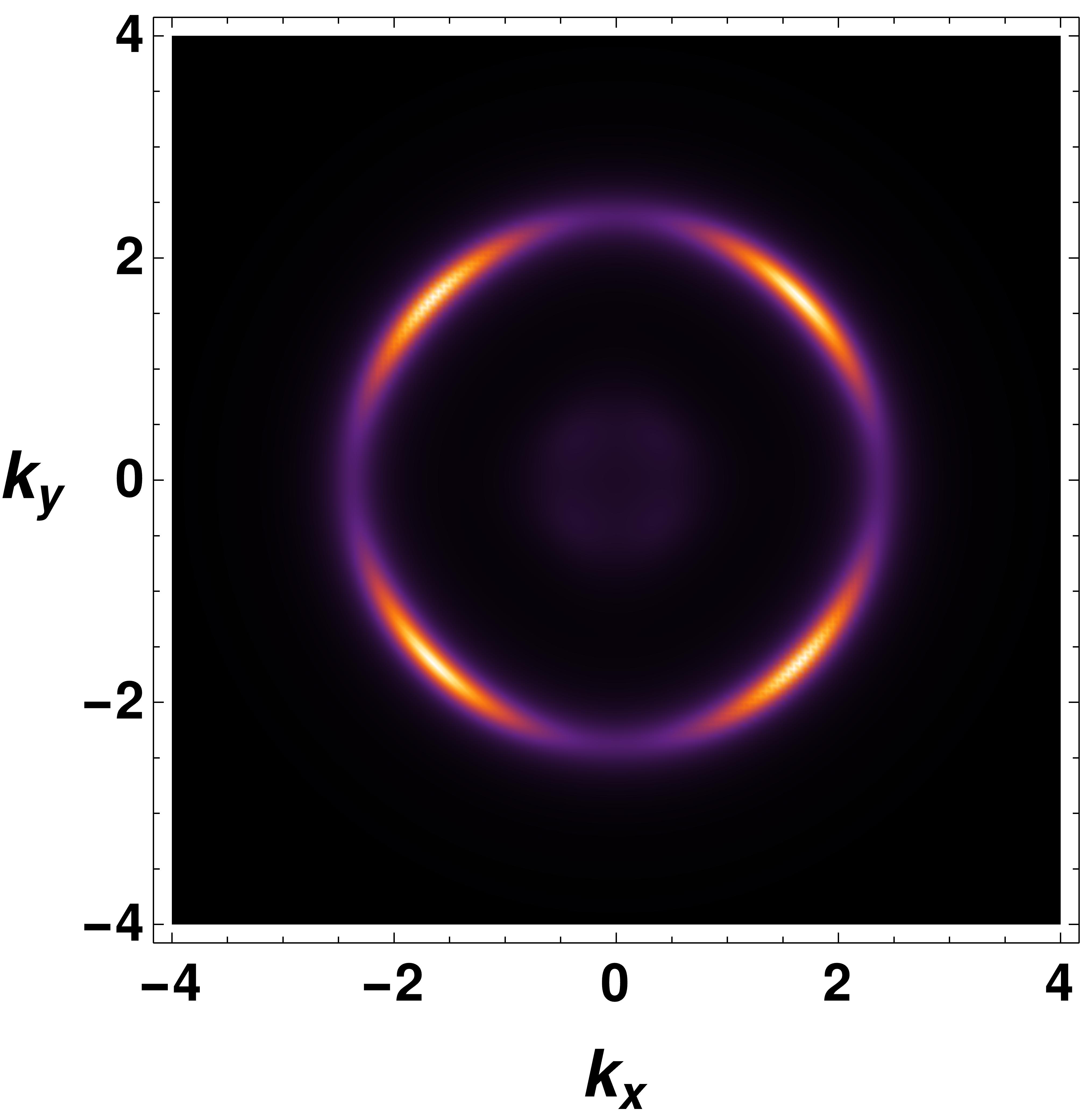}
		\caption{$\eta =0.5$}
	\end{subfigure}
	\hfil
	\begin{subfigure}[b]{0.3\textwidth}
		\centering
		\includegraphics[scale=0.23]{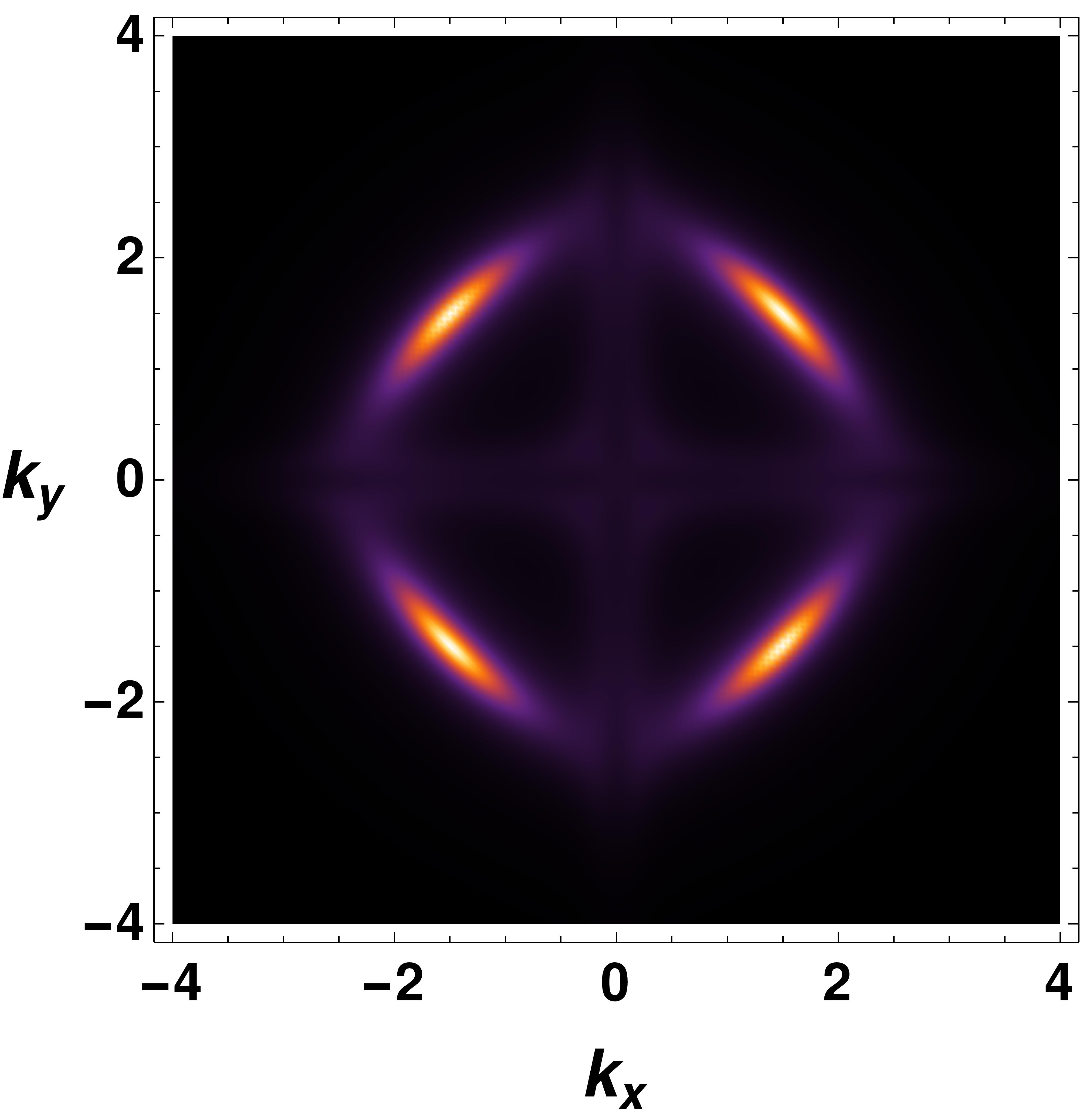}
		\caption{$\eta =2$}
	\end{subfigure}
	\hfil
	\begin{subfigure}[b]{0.3\textwidth}
		\centering
		\includegraphics[scale=0.225]{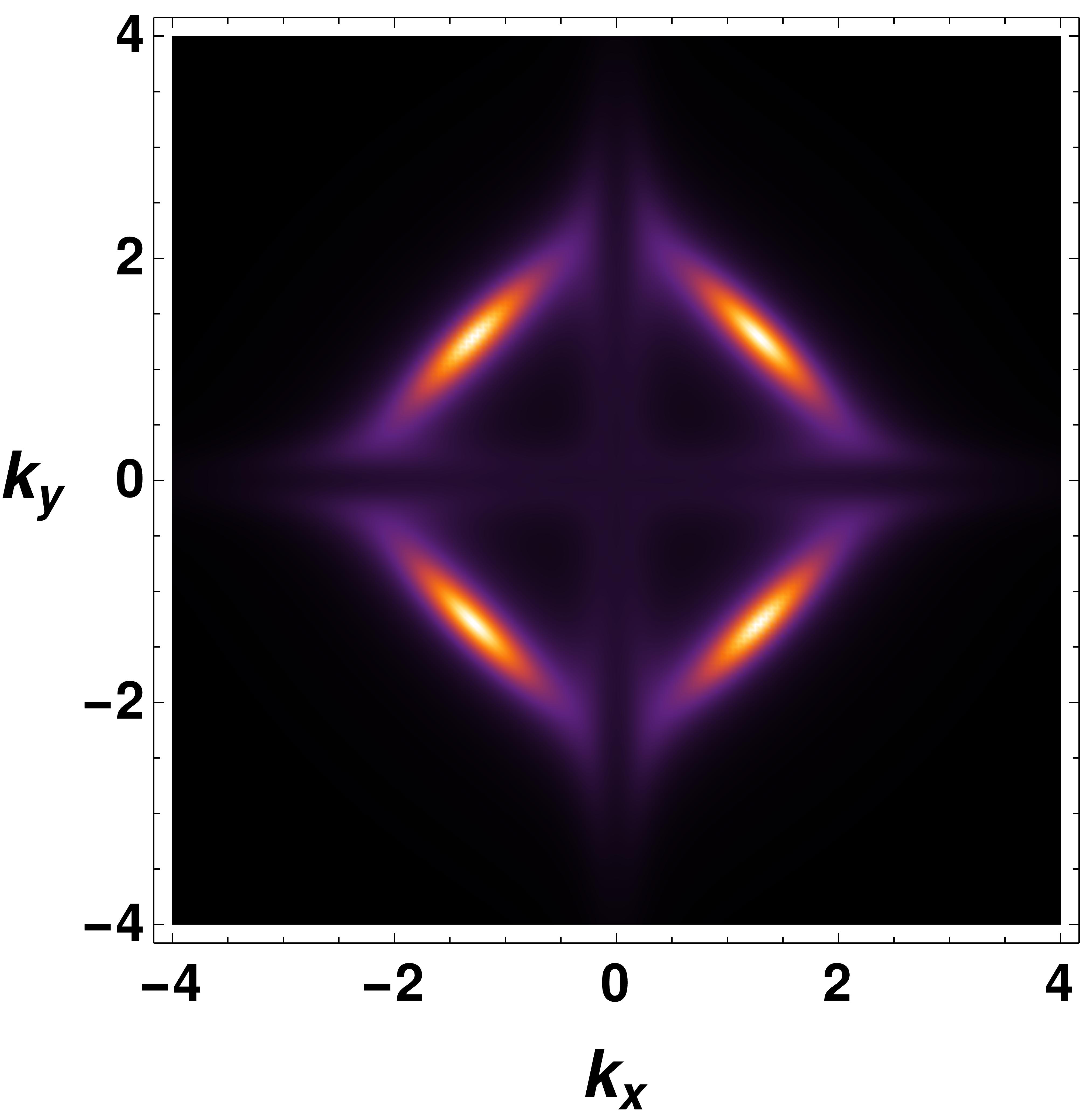}
		\caption{$\eta =5$}
	\end{subfigure}
	\caption{Evolution of spectral function with the coupling strength at $T=0.125 T_c$}
	\label{FigSF3eta}
\end{figure}

\noindent  By considering the tensor field interaction with fermions, we derive the $d$-wave fermionic spectral function, as depicted in Figure \ref{FigSF2}, for the $d_{x^2-y^2}$-orbital symmetry with the coupling constant $\eta=1$ at $T=0.125T_c$.
In the plot of $\omega$ versus $k_x=k_y=k$ (at the $45^{\circ}$ angle), no gap is observed, while a non-zero fermionic gap is evident in the $\omega$ versus $k_x$ plot. This arises due to the order parameter being zero at $45^{\circ}$ and maximal at $0^{\circ}$ in momentum space. Consequently, the $k_x$-versus-$k_y$ plot displays a Fermi arc along the $45^{\circ}$ angle in momentum space for the $d_{x^2-y^2}$ symmetry. The spectral function's band represents the particle and hole bands, where the fermionic gap is defined as the gap in $\omega$ within these two bands. For $d_{xy}$ symmetry, the position of the Fermi arc is shifted by approximately $45^{\circ}$ from that of $d_{x^2-y^2}$. Moreover, a mixture of both symmetries leads to an additional rotation of the Fermi arc by approximately $22.5^{\circ}$ compared to $d_{x^2-y^2}$, as demonstrated in equation (\ref{condval}) and Figures \ref{figgapm}, \ref{comdx2dxyb}. Therefore, we can confirm that in this holographic setup, the corrected angle-dependent order parameter is $B_{\rho\rho}$ rather than $B_{xx}$ or $B_{xy}$.
 \\
\noindent Now we provide a comparative analysis of the fermionic spectral function between the probe limit scenario and the case with backreaction, depicted in Figure \ref{FigSF3}, considering $\eta=1$ at $T=0.125T_c$. This illustration vividly demonstrates that the presence of backreaction exerts a substantial influence on the fermionic spectral function. Notably, the introduction of backreaction aids in nullifying the non-zero values of the spectral function inside region. 
\\
\noindent We also aim to explore the impact of the temperature-to-chemical-potential ratio on the fermionic spectral function. We have investigated the $\omega$-gap manifested in the fermionic spectral function, which directly correlates with the magnitude of the order parameter. 
Through the examination of bosonic configuration, we know that the value of the order parameter decreases as the temperature rises, eventually reaching zero at the critical temperature $T_c$ and beyond. Correspondingly, the fermionic gap decreases as the temperature increases, eventually vanishing at $T_c$, as illustrated in Figure \ref{FigSF3t}. 
\noindent Figure \ref{FigSF3eta} shows the influence of the tensor interaction strength on the fermionic gap. It is apparent that the Fermi arc deform little bit for higher value of coupling strength. The gap amplifies in magnitude as the coupling strength increases. The ARPES data \cite{RevModPhys.93.025006} shows that the structure of Fermi arc depends on the doping strength. Therefore, the coupling strength can be related with the doping parameter in the boundary theory. 
\\
To know the role of two flavor fermions in this set-up, which is related to sublattice symmetry in real materials, we have explored two flavor fermions with two different condensate in the Appendix \ref{app:first}. We have found that two fermion flavors with condensates of two different tensor fields lead to higher orbital spectral function.

\section{Discussion}
In this paper, we have done a comprehensive examination of the fermionic spectral function under the influence of a fully backreacted tensor condensation. 
We have determined the critical temperature to be $T_c=0.02\mu$ for the scaling dimension three and obtained all backreacted field configurations below this critical temperature.
By employing these field configurations along with tensor interactions of different orbital symmetries, specifically $d_{x^2-y^2}$, $d_{xy}$, and a combination of $d_{x^2-y^2}$ and $d_{xy}$, we have conducted numerical investigations into the fermionic spectral function. This has been accomplished by solving the flow equation for the bulk Green function. Our analysis has unveiled the presence of a $d$-wave Fermi arc in the presence of the tensor field. \\

\noindent The momentum dependent order parameter, denoted as $\Delta_k$, is identified as  angle-dependent tensor field component $B_{\rho\rho}$ in momentum space. Comparing the  the momentum-dependence of the order parameter (Fig.\ref{figgapm}) with that of the fermion spectral function  (Fig.\ref{comdx2dxyb}), we confirm the correct order parameter for $d$-wave holographic superconductors. The $d_{x^2-y^2}+d_{xy}$-wave condensation rotates the Fermi arc position which confirms the $d$-wave order parameter (\ref{condval}).
We compare the spectral function in the probe limit case with the backreacted case. The $d+id$-wave condensation leads $s$-wave fermionic gap which exactly matches with previous findings \cite{Chen:2011ny}. Similary, the $p+ip$-wave condensation creates the $s$-wave fermionic spectral function. \\


\noindent The higher value of the coupling constant decreases the convexity of the Fermi arc.
 We have observed that as the temperature increases, the condensation value decreases, leading to a reduction in the fermionic gap. When the system reaches its critical temperature, it undergoes a transition to the normal phase. This transition is characterized by the emergence of a Fermi surface in the spectral function, attributed to the closure of the superconducting gap. The spectral function closely aligns with the experimental findings for high $T_c$ superconductors mentioned in \cite{RevModPhys.93.025006}. Furthermore, we have analyzed the role of two-flavor fermions in the presence of a vector field and a tensor field, resulting in a higher orbital symmetric fermionic spectral function. The vector condensate in the $p_x$ and $p_y$ directions, combined with two-flavor fermions, leads to a $d$-wave fermionic spectral function. Moreover, the combination of two $d$-wave symmetries ($d_{x^2-y^2}$ and $d_{xy}$) with two-flavor fermions results in a $g$-wave like fermionic spectral function. This may have implications for the discovery of superconductivity with higher orbital symmetry. 
 For the study of the unconventional superconductors within a holographic framework, the constructing the superconducting dome through an examination of the spectral function is  an intriguing avenue, which is our future direction.

\appendix
\section{Two flavour fermions: Higher orbital spectral function}
\label{app:first}
Here, we delve into the investigation of two-flavor fermions within the context of vector and tensor condensation. The underlying motivation for exploring this scenario is to know whether the combination of two condensates with two-flavor fermions results in a spectral function exhibiting higher or lower orbital symmetry.

\subsection{With tensor field}
\begin{figure}[h!]
	\centering		
	\begin{subfigure}[b]{0.45\textwidth}
		\centering
		\includegraphics[scale=0.23]{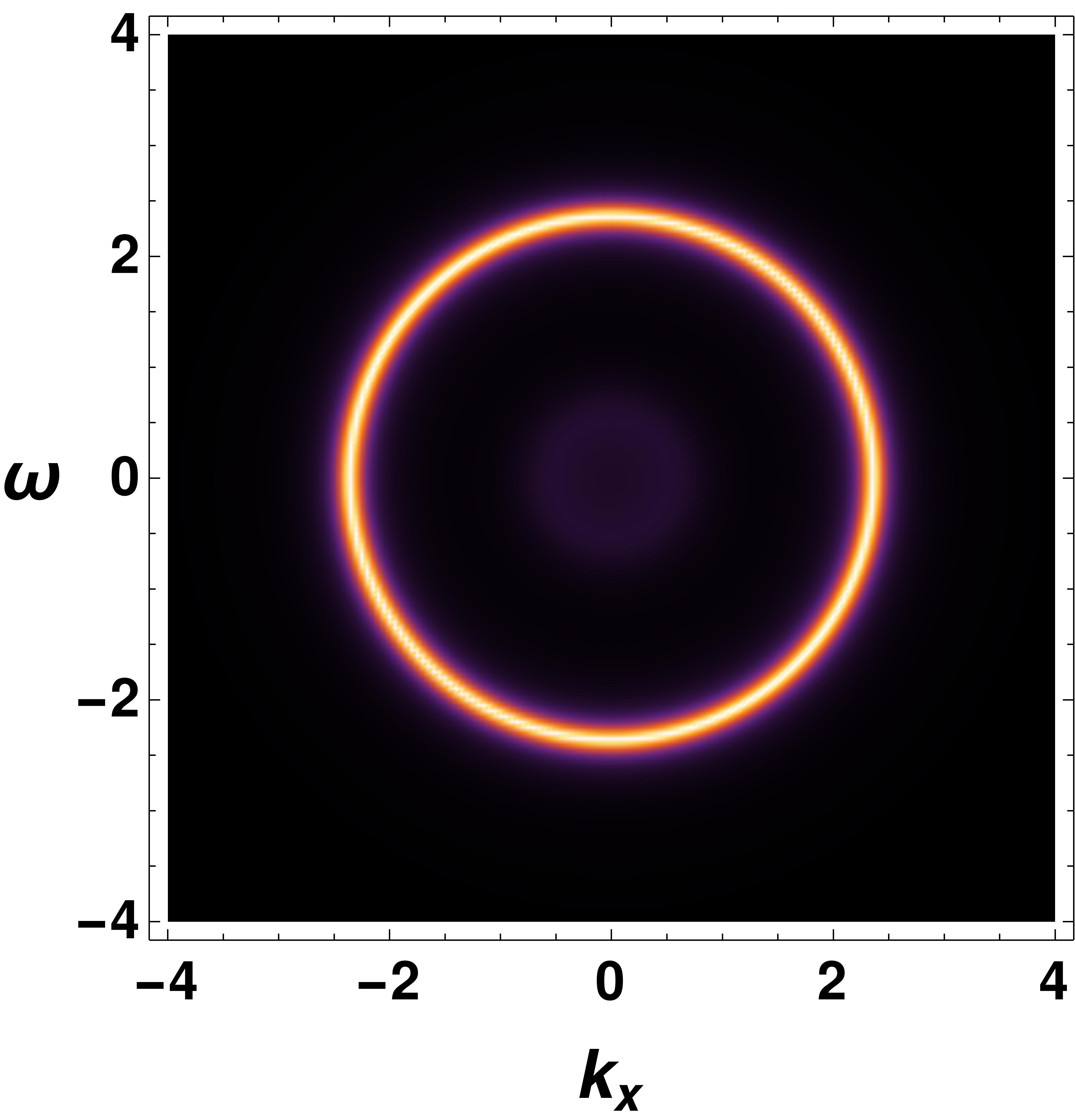}
		\caption{ No interaction $\eta=0$}
	\end{subfigure}
	\hfil
	\begin{subfigure}[b]{0.45\textwidth}
		\centering
		\includegraphics[scale=0.23]{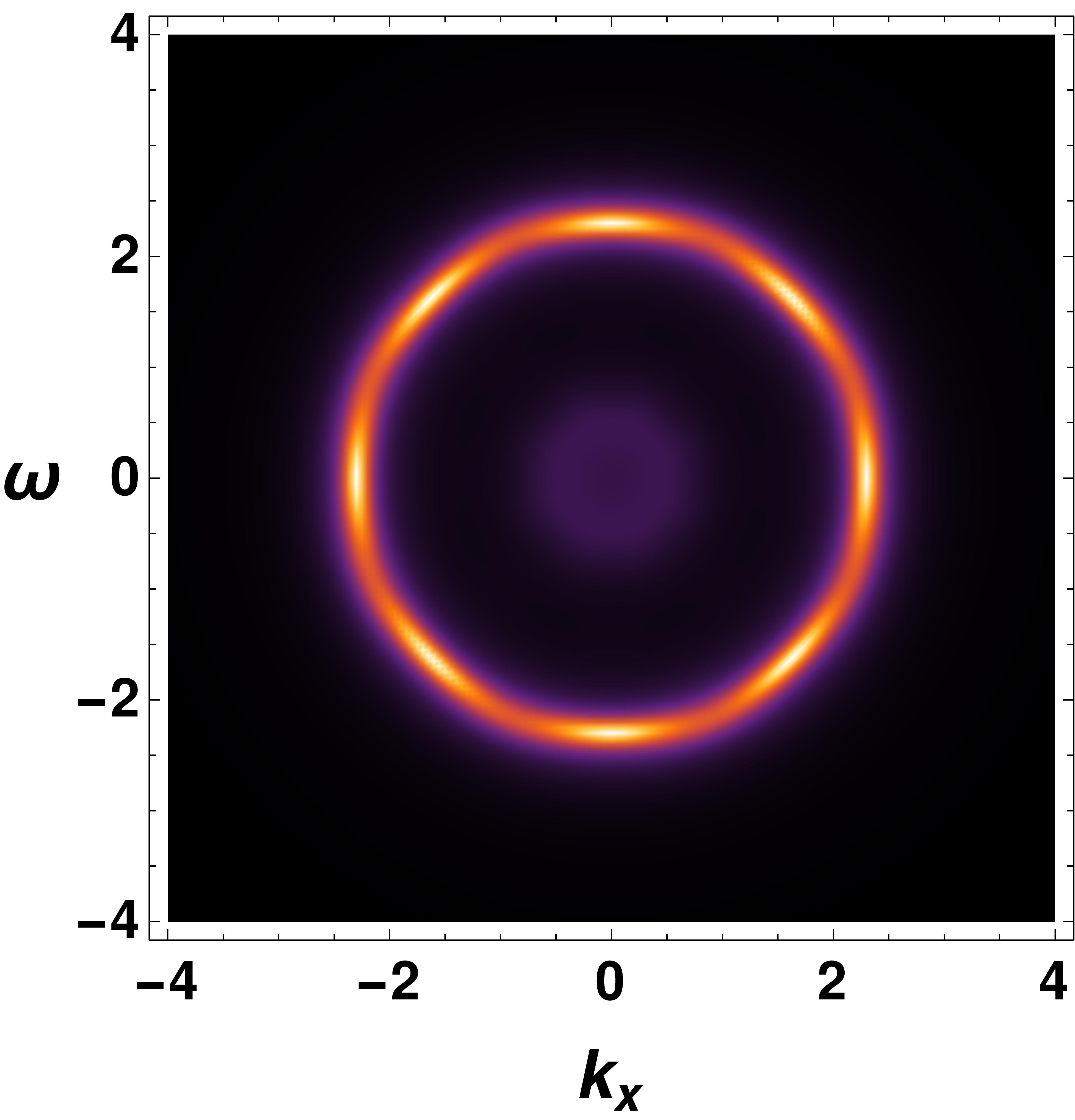}
		\caption{With both interaction $\eta=1 $}
	\end{subfigure}
	\caption{The $g$-wave spectral function from two flavour fermions with two $d$-waves ($d_{x^2-y^2}$ and $d_{xy}$) condensate at $T=0.125 T_c$. (a) Fermi surface: when no interaction or one of interaction turn off $(\alpha=0~ \text{or}~ \beta=0)$. (b) The $g$-wave: when both interaction turn on $\eta=1, \alpha=\frac{1}{\sqrt{2}}, \beta=\frac{1}{\sqrt{2}}$ which means that flavour one is coupling with $d_{x^2-y^2}$ condensate and another flavour is coupling with $d_{xy}$ condensate.}
	\label{figfd}
\end{figure}

For tensor field, we have considered $d_{x^2-y^2}$ condensate with flavour one fermion and $d_{xy}$ condensate with another flavour fermion. The corresponding interaction Lagrangian density is
\begin{eqnarray}
	\mathcal{L}_{int} &=& \sum_{i=1,2} \left[ \eta^{*} B^{*(i)}_{\mu\nu} \bar{\psi_c}^{(i)} \Gamma^{\mu}D^{\nu}\psi^{(i)}- \eta \bar{\psi}^{(i)}\Gamma^{\mu} D^{\nu}(B^{(i)}_{\mu\nu} \psi^{(i)}_{c})\right]
\end{eqnarray}
where $i$ is the flavour index, and $B^{(1)}=\frac{\alpha\phi(z)}{\sqrt{2}z^2}\left[dx^2-dy^2\right]$ has $d_{x^2-y^2}$ symmetry and  $B^{(2)}=2\frac{\beta\phi(z)}{\sqrt{2}z^2} dxdy$ has $d_{xy}$ symmetry. We can easily promot our previous calculation (one flavour) to two flavour by promoting the source and condensation in following form
\begin{eqnarray}
	\xi^{(C)}= \begin{pmatrix} 
		\Psi^{(1)}_{-} \\
		\Psi^{(2)}_{-} \\
		\Psi^{(1)}_{c+} \\
		\Psi^{(2)}_{c+}
	\end{pmatrix},  ~~~~~~~\text{and}~~ \xi^{(S)}= \begin{pmatrix} 
		\Psi^{(1)}_{+} \\
		\Psi^{(2)}_{+} \\
		\Psi^{(1)}_{c-} \\
		\Psi^{(2)}_{c-} 
	\end{pmatrix}.
\end{eqnarray}
The above expression is defined from following boundary action
\begin{eqnarray}
	S_{bdy}= i \int d^3x  \sqrt{-h} \sum_{i=1,2}\left[ \bar{\psi}^{(i)}\psi^{(i)}+ \bar{\psi_c}^{(i)}\psi^{(i)}_c\right] ~.
\end{eqnarray}
Using above setup, one can find a similar flow equation 
\begin{eqnarray}
	\partial_z \mathbb{G}(z) + \tilde{\Gamma} \mathbb{M}_3 \tilde{\Gamma} \mathbb{G}(z) - \mathbb{G}(z) \mathbb{M}_1 -\mathbb{G}(z) \mathbb{M}_2\tilde{\Gamma} \mathbb{G}(z)+ \tilde{\Gamma} \mathbb{M}_4 =0
\end{eqnarray}
where $\mathbb{G}(z)$ is $8\times 8$ bulk Green function matrix and all $\mathbb{M}_i$ are given by
\begin{eqnarray}
	\mathbb{M}_1= \begin{pmatrix}
		\mathbb{N}_1 & \mathbb{P}_1 \\
		-\mathbb{P}_1 & - \mathbb{N}_1
	\end{pmatrix}, 		~~~\mathbb{M}_2= \begin{pmatrix}
		\mathbb{N}_2(q) & 0 \\
		0 & \mathbb{N}_2 (-q)
	\end{pmatrix}, 	~~~\mathbb{M}_3= -\mathbb{M}_1 ~~~~ \mathbb{M}_4= -\mathbb{M}_2 ~~~~~~~
\end{eqnarray}
where
\begin{eqnarray}
	\mathbb{N}_1 = - \frac{m_f}{z \sqrt{f(z)}}\boldsymbol{1}_{4\times 4}, ~~~~	\mathbb{P}_1 =\frac{\sqrt{2}\eta \phi(z)}{\sqrt{f(z)}}\begin{pmatrix}
		- \alpha k_y & \alpha  k_x  & 0 & 0\\
		\alpha k_x  & \alpha  k_y & 0 & 0 \\
		0 & 0 & \beta k_x & \beta k_y \\
		0 & 0 & \beta k_y &  - \beta k_x
	\end{pmatrix}   \nonumber \\
	\mathbb{N}_2 (q) = \frac{i}{\sqrt{f(z)}} \begin{pmatrix}
		\mathbb{P}_2 (q) & 0 \\
		0 & \mathbb{P}_2 (q)
	\end{pmatrix}	~~~ \text{and} ~~~\mathbb{P}_2 (q) = \begin{pmatrix}
		k_y  & k_x-\frac{(\omega+ q A_t(z))}{\sqrt{f(z) g(z)}} \\
		k_x+\frac{(\omega+ q A_t(z))}{\sqrt{f(z) g(z)}} & -k_y 
	\end{pmatrix} ~.~~~
\end{eqnarray}
The horizon value of the bulk green function is $\mathbb{G} (z_h) = i \bold{1}_{8\times 8}$. With this, we numerically calculated the spectral function in presence of backreaction which shows $g$-wave like fermionic gap in the Figure \ref{figfd}. 


	\subsection{With vector field}
\begin{figure}[h!]
		\centering		
		\begin{subfigure}[b]{0.24\textwidth}
			\centering
			\includegraphics[scale=0.21]{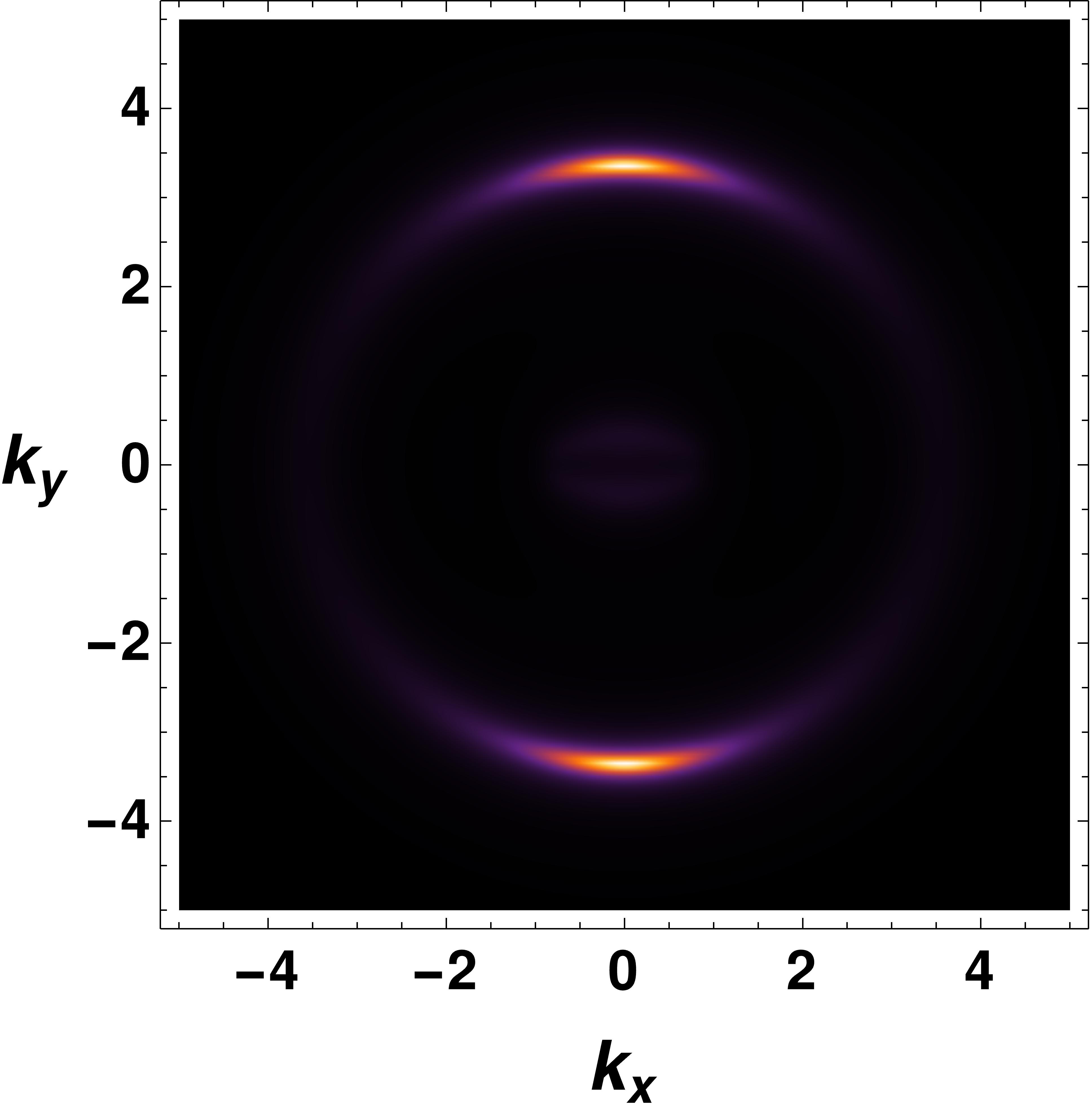}
			\caption{$p_x$-wave:\\
				$~~~~~~(\tilde{\alpha},\tilde{\beta})=(1,0)$}
		\end{subfigure}
		\hfil
		\begin{subfigure}[b]{0.24\textwidth}
			\centering
			\includegraphics[scale=0.21]{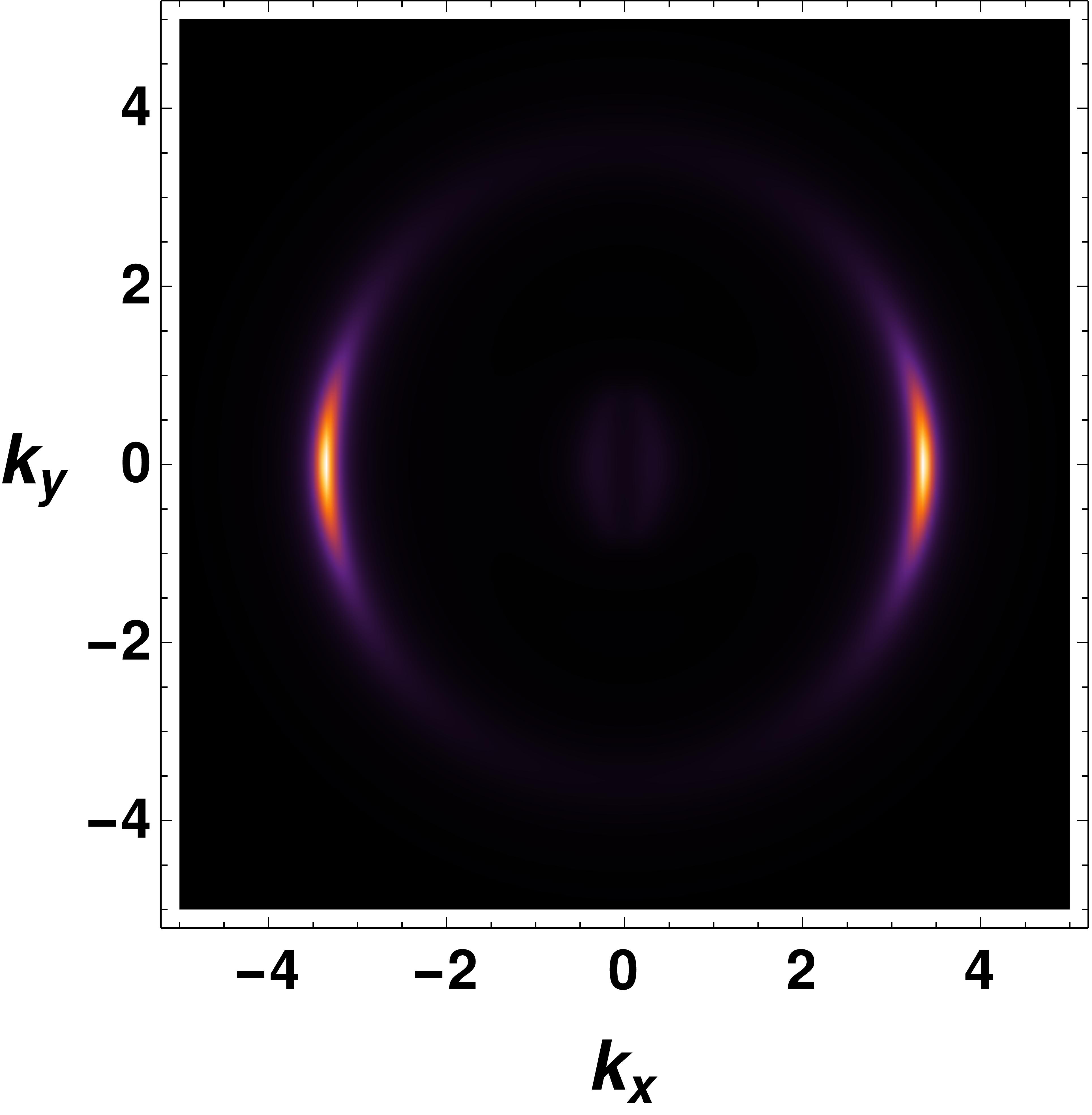}
			\caption{$p_y$-wave: \\
				$~~~~~~(\tilde{\alpha},\tilde{\beta})=(0,1)$}
		\end{subfigure}
		\hfil
		\begin{subfigure}[b]{0.24\textwidth}
			\centering
			\includegraphics[scale=0.21]{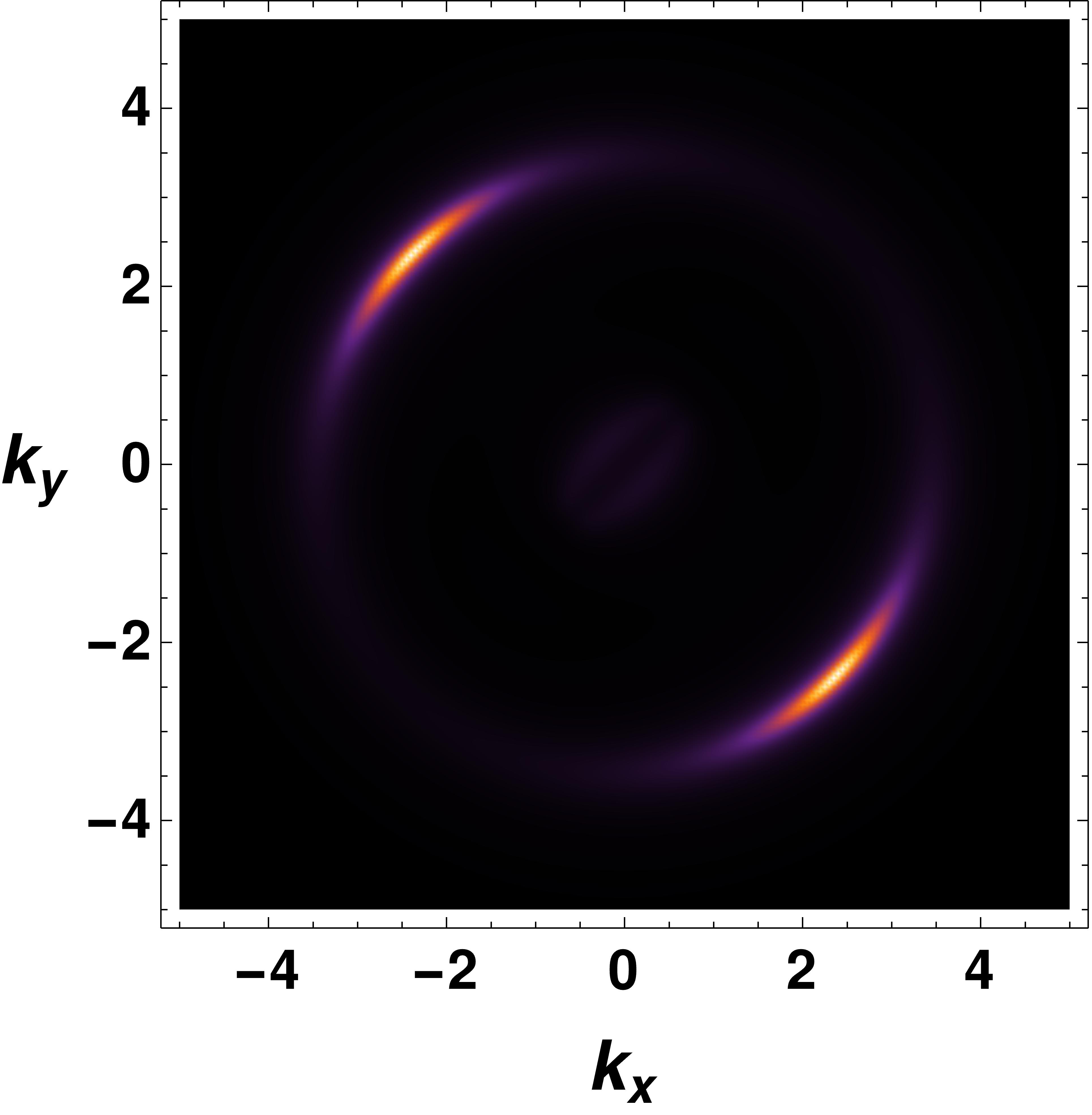}
			\caption{$p_x+p_y$-wave: \\
				$~~~~~~(\tilde{\alpha},\tilde{\beta})=(\frac{1}{\sqrt{2}},\frac{1}{\sqrt{2}})$}
		\end{subfigure}
		\hfil
		\begin{subfigure}[b]{0.24\textwidth}
			\centering
			\includegraphics[scale=0.20]{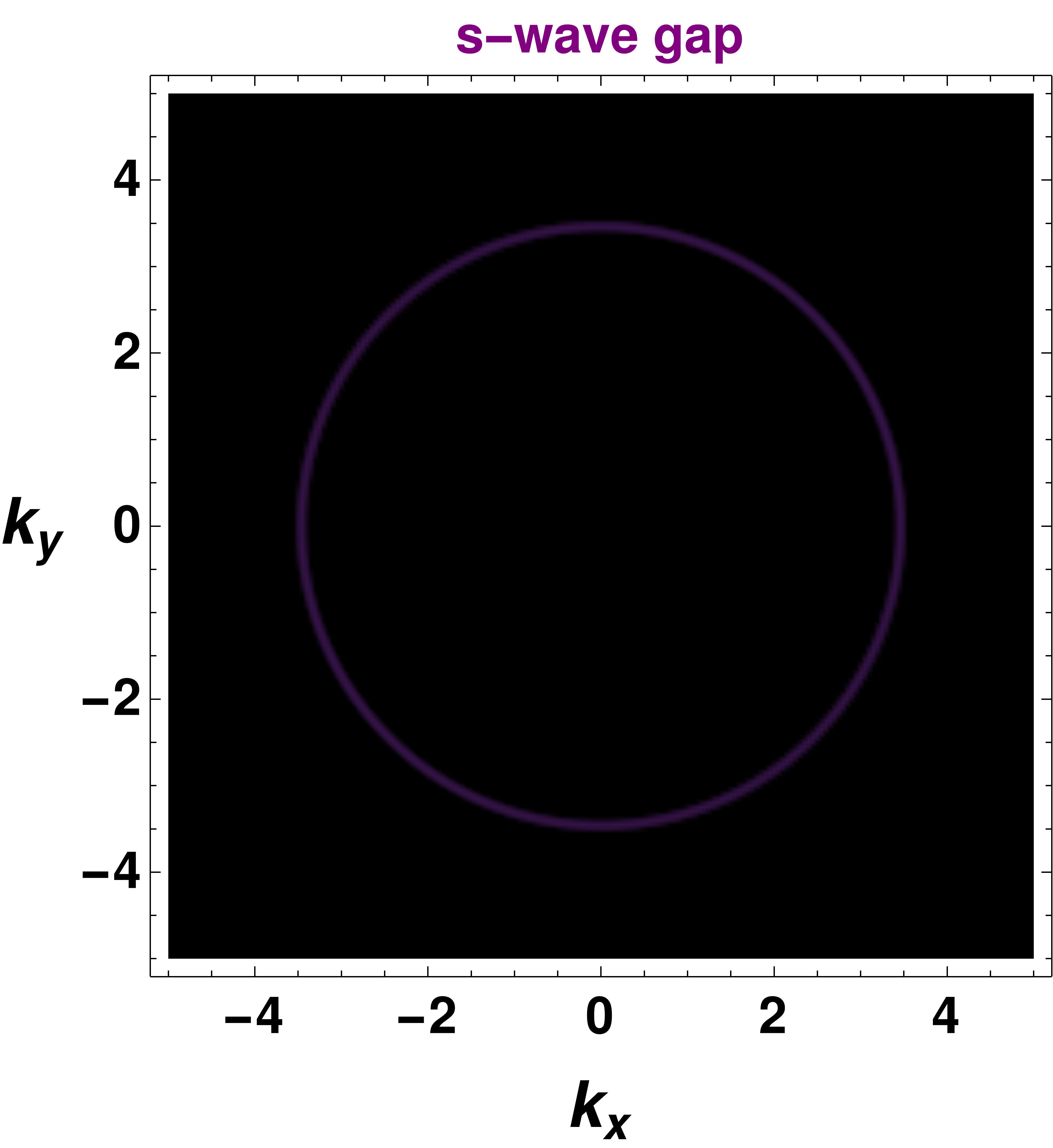}
			\caption{$p_x+ip_y$-wave: \\
				$~~~~~~(\tilde{\alpha},\tilde{\beta})=(\frac{1}{\sqrt{2}},\frac{i}{\sqrt{2}})$}
		\end{subfigure}
		\caption{One flavour fermionic spectral function with vector condensate at $T=0.233T_c$. The plot (d) shows a $s$-wave fermionic spectral function from $p_x+ip_y$-wave order parameter.}
		\label{dwavep1}
	\end{figure}

	Here, we start with one flavour fermion with $p_x$-wave and $p_y$-wave vector condensate. We will show that the $p_x+ p_y$-wave rotates the Fermi arc position in momentum space. The Lagrangian for vector condensate reads 
	\begin{eqnarray}
		S_b = \int d^4x \sqrt{-g} \left[\frac{1}{2\kappa^2} \left(R -2 \Lambda\right) + \mathcal{L}_v \right]
	\end{eqnarray}
	where the matter Lagrangian density 
	\begin{eqnarray}
		\mathcal{L}_{v}= -\frac{1}{4}F_{\mu\nu}F^{\mu\nu}-\frac{1}{2}V^{\dagger}_{\mu\nu}V^{\mu\nu}- m^2 V^{\dagger}_{\mu}V^{\mu} ~.
		\label{pwavev0}
	\end{eqnarray}
	The covariant derivative of the vector field is defined as $V_{\mu\nu}=\partial_{\mu}V_{\nu}-\partial_{\nu}V_{\mu}-i q_v A_{\mu}V_{\nu}+i q_v A_{\nu}V_{\mu}$.
	With vector field ansatz $V=\phi_{p}(z)\left[\tilde{\alpha} dx + \tilde{\beta} dy\right]$, we obtain
	\begin{eqnarray}
		\label{pbeom1a}
		g'+\frac{2\kappa^2 z^3}{L^2} p^2_m  \left[ g \phi_{p}^{\prime 2} + \frac{q_v^2 A_t^2 \phi_{p}^2}{f^2} \right] &=& 0  \\
		\label{pbeom1b}
		f'-\frac{3 }{z}f +\frac{3}{z} + \frac{g^{\prime}}{2g} f - \kappa^2 z \left[m^2 p^2_m \phi_p^2 +\frac{z^2 A^{\prime 2}_t}{2L^2 g} \right] &=& 0  \\
		\label{pbeom1c}
		A_t^{\prime\prime}-\frac{g^{\prime}}{2g}A_t^{\prime}-\frac{2 q_v^2 \phi_{p}^2}{f} p^2_m A_t &=& 0  \\
		\label{pbeom1d}
		\phi_{p}^{\prime\prime}+ \left[\frac{f'}{f}+\frac{g'}{2g} \right]\phi_{p}^{\prime}+\left[\frac{q_v^2 A_t^2}{f^2 g}-\frac{m^2 L^2}{z^2 f}\right]\phi_{p} &=& 0 ~~~
	\end{eqnarray}
	where $p^2_m=|\tilde{\alpha}|^2+ |\tilde{\beta}|^2$.
	Using horizon and boundary condition of the vector field \cite{Ghorai:2023wpu}, we solve all the fours fields for given value of $\frac{T}{\mu}$. The interaction between vector field and fermion is given by $\mathcal{L}_{int}=  \bar{\psi} V_{\mu} \Gamma^{\mu} \psi_c + h.c.$. Using this interaction, we find the flow equation 
	\begin{eqnarray}
		\partial_z \mathbb{G}(z) + \tilde{\Gamma} \mathbb{M}_3 \tilde{\Gamma} \mathbb{G}(z) - \mathbb{G}(z) \mathbb{M}_1 -\mathbb{G}(z) \mathbb{M}_2\tilde{\Gamma} \mathbb{G}(z)+ \tilde{\Gamma} \mathbb{M}_4 =0
		\label{equflowmm}
	\end{eqnarray}
	where the matrix $\mathbb{M}_i$ is given by
	\begin{eqnarray}
		\mathbb{M}_1= \begin{pmatrix}
			\mathbb{N}_1 & \mathbb{P}_1 \\
			\mathbb{P}_1 & - \mathbb{N}_1
		\end{pmatrix}, 		~~~\mathbb{M}_2= \begin{pmatrix}
			\mathbb{N}_2(q) & 0 \\
			0 & \mathbb{N}_2 (-q)
		\end{pmatrix}, 	~~~\mathbb{M}_3= -\mathbb{M}_1 ~~~~ \mathbb{M}_4= -\mathbb{M}_2 ~~~~~~~
	\end{eqnarray}
	where
	\begin{eqnarray}
		\mathbb{N}_1 = - \frac{m_f}{z \sqrt{f(z)}}\boldsymbol{1}_{2\times 2}, ~~~~	\mathbb{P}_1 =\frac{i \phi_p(z)}{\sqrt{f(z)}}\left[\tilde{\alpha} \begin{pmatrix}
			0 &  -1  \\
			-1  &  0
		\end{pmatrix} + \tilde{\beta} \begin{pmatrix}
			-1 &  0 \\
			0 &  1
		\end{pmatrix} \right]  \nonumber \\
		\mathbb{N}_2 (q) = \frac{i}{\sqrt{f(z)}} \begin{pmatrix}
			k_y  & k_x-\frac{(\omega+ q A_t(z))}{\sqrt{f(z) g(z)}}\\
			k_x+\frac{(\omega+ q A_t(z))}{\sqrt{f(z) g(z)}} & -k_y
		\end{pmatrix}	~.~~~
	\end{eqnarray}
	The spectral function at $T=0.233 T_c$ is shown in Figure \ref{dwavep1} for $\Delta=2 (m^2=0)$ and $m_f=0$. The $p_x+ip_y$-wave rotates the Fermi arc position which confirms again that the $p$-wave order paramater \cite{Ghorai:2023wpu} is $V_{\rho}=\phi_{p}(z)\left[\tilde{\alpha}\cos\theta+\tilde{\beta}\sin\theta \right]$. An $s$-wave fermionic spectral function can be derived from the $p_x+ip_y$-wave order parameter, similar to how we obtained it from the $d_{x^2-y^2}+id_{xy}$-wave order parameter.
	\subsubsection{Two flavour fermions with vector field} 
\begin{figure}[h!]
		\centering		
		\includegraphics[scale=0.2]{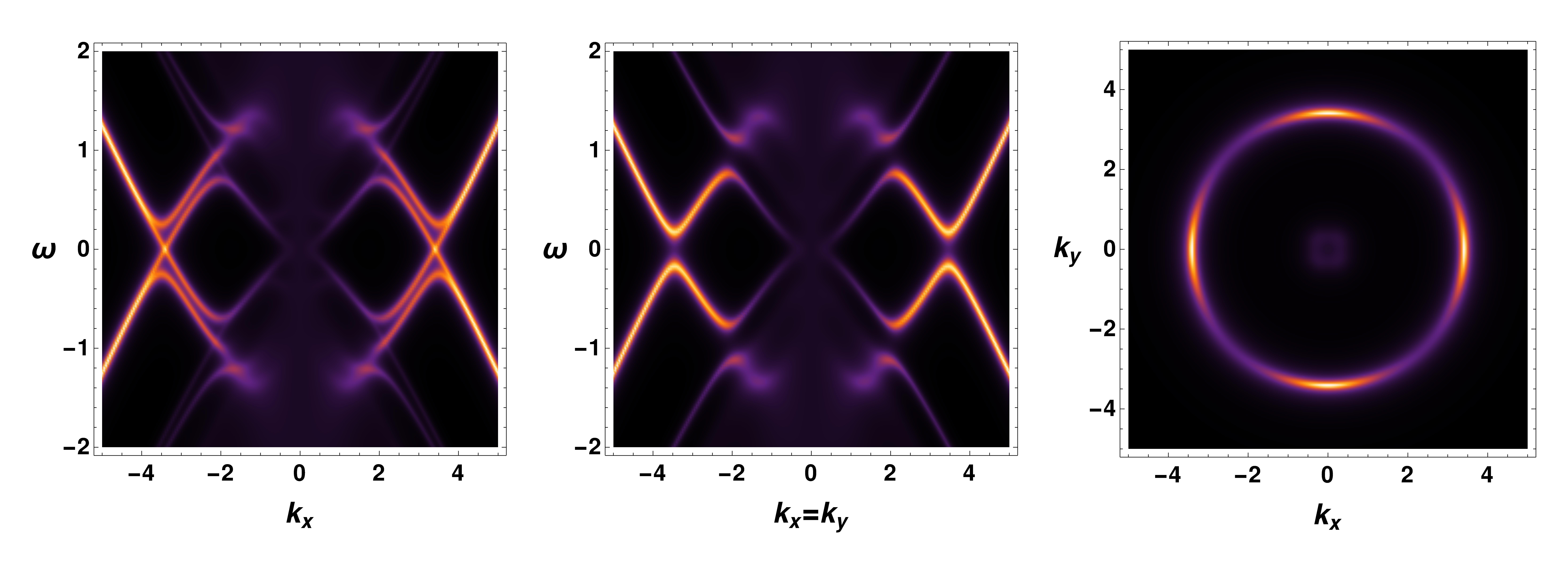}
		\caption{Spectral function in the presence of the interaction of one flavour fermion with $p_x$-wave condensate and the interaction of another flavour fermion with $p_{y}$-wave condensate at $T=0.233T_c$ shows a $d$-wave spectral function. }
	\label{figfp}
\end{figure}

The interaction term of two flavour fermion with $p_x$ and $p_y$ vector condensate can be expressed in following form
	\begin{eqnarray}
		\mathcal{L}_{int} &=& \sum_{i=1,2} \left[
		\bar{\psi}^{(i)} V_{\mu}^{(i)}\Gamma^{\mu} \psi_c^{(i)} + h.c. \right]
	\end{eqnarray}
	where $V^{(1)}=\tilde{\alpha} \phi_p(z) dx$ and $V^{(2)}=\tilde{\beta} \phi_p(z) dy$. The flow equation becomes
	\begin{eqnarray}
		\partial_z \mathbb{G}(z) + \tilde{\Gamma} \mathbb{M}_3 \tilde{\Gamma} \mathbb{G}(z) - \mathbb{G}(z) \mathbb{M}_1 -\mathbb{G}(z) \mathbb{M}_2\tilde{\Gamma} \mathbb{G}(z)+ \tilde{\Gamma} \mathbb{M}_4 =0 
	\end{eqnarray}
	where $\mathbb{G}(z)$ is $8\times 8$ matrix and the matrix $\mathbb{M}_i$ is given by
	\begin{eqnarray}
		\mathbb{M}_1= \begin{pmatrix}
			\mathbb{N}_1 & \mathbb{P}_1 \\
			\mathbb{P}_1 & - \mathbb{N}_1
		\end{pmatrix}, 		~~~\mathbb{M}_2= \begin{pmatrix}
			\mathbb{N}_2(q) & 0 \\
			0 & \mathbb{N}_2 (-q)
		\end{pmatrix}, 	~~~\mathbb{M}_3= -\mathbb{M}_1 ~~~~ \mathbb{M}_4= -\mathbb{M}_2 ~.~~~~~
	\end{eqnarray}
	All $4\times 4$ matrix in the above expression are given by
	\begin{eqnarray}
		\mathbb{N}_1 = - \frac{m_f}{z \sqrt{f(z)}}\boldsymbol{1}_{4\times 4} &,& ~~~~	\mathbb{P}_1 =\frac{i \phi_p(z)}{\sqrt{f(z)}}\begin{pmatrix}
			0 & -\tilde{\alpha}   & 0 & 0\\
			-\tilde{\alpha}  & 0 & 0 & 0 \\
			0 & 0 & -\tilde{\beta} & 0 \\
			0 & 0 & 0 &  \tilde{\beta}
		\end{pmatrix}   \nonumber \\
		\mathbb{N}_2 (q) = \frac{i}{\sqrt{f(z)}} \begin{pmatrix}
			\mathbb{P}_2 (q) & 0 \\
			0 & \mathbb{P}_2 (q)
		\end{pmatrix}	~ &,& ~~~\mathbb{P}_2 (q) = \begin{pmatrix}
			k_y  & k_x-\frac{(\omega+ q A_t(z))}{\sqrt{f(z) g(z)}} \\
			k_x+\frac{(\omega+ q A_t(z))}{\sqrt{f(z) g(z)}} & -k_y 
		\end{pmatrix} ~.~~~~~
	\end{eqnarray}
	The resulting spectral function is given in Figure \ref{figfp}. This shows a clear $d$-wave fermionic spectral function from the interaction of flavour one with $p_x$-wave condensate and another flavour with $p_{y}$-wave condensate at $T=0.233T_c$.
Therefore, combining two condensates with two-flavor fermions gives a spectral function exhibiting higher orbital symmetry.

\acknowledgments
This work is supported by Mid-career Researcher Program through the National Research
Foundation of Korea grant No. NRF-2021R1A2B5B02002603, RS-2023-00218998 and NRF-2022H1D3A3A01077468. 
We thank the APCTP for the hospitality during the focus program, where part of this work
was discussed.

\bibliographystyle{jhep}
\bibliography{refpapers.bib}
\end{document}